\documentclass[fleqn,usenatbib]{mnras}
\usepackage[utf8]{inputenc}
\usepackage{mathrsfs}
\usepackage{natbib}
\usepackage{amsmath,amssymb}
\usepackage{amsfonts}
\usepackage{graphicx}
\usepackage{hyperref}
\usepackage[usenames,dvipsnames]{xcolor}
\usepackage{mathrsfs}

\makeatletter

\usepackage{mathrsfs}

\newcommand{\set}[1]{\left\{#1\right\}}

\newcommand{\eps}{\varepsilon}

\newcommand{\zhat}{\mathbf{\hat{z}}}

\newcommand{\xhat}{\mathbf{\hat{x}}}

\makeatother

\title[Quasi-Hierarchical Triples]{Dynamical Evolution of Quasi-Hierarchical Triples}

\author[Ginat, Stegmann and Samsing]{Yonadav Barry Ginat,$^{1,2 }$\thanks{E-mail: yb.ginat@physics.ox.ac.uk} Jakob Stegmann$^{3}$ and
Johan Samsing$^{4}$
\\
$^{1}$Rudolf Peierls Centre for Theoretical Physics, University of Oxford, Parks Road, Oxford, OX1 3PU, United Kingdom \\ 
$^{2}$New College, Holywell Street, Oxford, OX1 3BN, United Kingdom \\ 
$^{3}$Max-Planck-Institut f\"ur Astrophysik, Karl-Schwarzschild-Stra\ss e~1, 85748 Garching bei M\"{u}nchen, Germany \\
$^{4}$Niels Bohr Institute, Copenhagen University, Blegdamsvej 17, 2100 Copenhagen, Denmark
}

\date{Accepted XXX. Received YYY; in original form ZZZ}

\pubyear{2026}
\begin{document}
\label{firstpage}
\pagerange{\pageref{firstpage}--\pageref{lastpage}}
\maketitle

\begin{abstract}
  We study the gravitational dynamics of quasi-hierarchical triple systems, where the outer orbital period is significantly longer than the inner one, but the outer orbit is extremely eccentric, rendering the time at pericentre comparable to the inner period. Such systems are not amenable to the standard techniques of perturbation theory and orbit-averaging. Modelling the evolution of these triples as a sequence of impulses at the outer pericentre, we show, by comparing with direct three-body integrations, that such triples lend themselves to a description as an analytical map between subsequent outer pericentre passages. This map exhibits secular oscillations, going beyond the von Zeipel--Lidov--Kozai mechanism. We show that the time to coalescence due to gravitational waves in such systems is modified. We then study the long-term evolution under this map, which lead to a random-walk-like behaviour of the inner eccentricity. While this behaviour is probably absent from isolated triples, it could exist in triples where the outer orbit is weakly coupled to a system with which it can exchange angular momentum, and we describe some properties of this random walk.
\end{abstract}



\section{Introduction}
\label{sec:introduction}

Triple systems are home to an immense range of dynamical and astrophysical phenomena \citep{ValtonenKarttunen2006,Perets2025}. They play an important role in the evolution of all many-body astrophysical systems: from planetary systems \citep{2015ARA&A..53..409W}, to massive stars \citep{Offner2023,Shariatetal2025}, stellar clusters \citep{Antonini2016,Martinez2020,2022MNRAS.511.1362T}, and galaxies \citep[e.g.,][]{Mareletal2012}.

Triple systems of similar component-masses are generally either hierarchical (there is a well-defined inner binary orbited by an outer star, with a clear separation of orbital time-scales) or non-hierarchical (the three stars are in an approximate energy equipartition) and unstable \citep{Perets2025}. Hierarchical triples, which can be stable over long time-scales, can be mathematically treated by means of perturbation theory \citep[e.g.,][]{Harrington1968,Ford2000,Arnoldetal2006,Katzetal2011,Naoz2013,Tremaine2023}. Indeed, such systems are usually studied either by direct three-body integrations, or by various analytical techniques, in a hierarchy of approximations (or coarse-graining operations), as one is concerned with the long-term state of the system, rather than the rapid temporal evolution. These include a single orbit-averaging over the inner binary's fast period, where the orbital elements of the inner binary evolve over time-scales comparable to the outer orbit, due to the tertiary's influences, or double orbit-averaging, where the system is averaged over both the inner and the outer orbit \cite[see][for reviews]{Naoz2016,Shevchenko2017}. These approximations are adequate when one is concerned with time-scales much longer that the outer period, and in the absence of strong resonances.\footnote{Note, that this is distinct from a double-averaging procedure of the inner orbit, where the inner pericentre's precession is averaged over---we do not consider such an averaging here.} An example of such a phenomenon is the famous von Zeipel--Lidov--Kozai effect \citep[ZLK;][]{Zeipel1910,Lidov1962,Kozai1962}.

Non-hierarchical, or `democratic', triples are notoriously chaotic and unstable \citep{Hut1993,Heinamakietal1999,MardlingAarseth2001,ValtonenKarttunen2006,Zhangetal2023,Tranietal2024b}, but ensembles of such systems---as all dense clusters are---may be modelled using statistical theories \citep{Heggie1975,Monaghan1976a,Monaghan1976b,ValtonenKarttunen2006,StoneLeigh2019,GinatPerets2021a,Kol2021}.

Hierarchical triples, even in the Galactic field, do not exist in isolation: the outer orbit can be wide, and is susceptible to external perturbations, which can affect it significantly \citep[e.g.,][]{MichaelyPerets2019,MichaelyPerets2020,Ravehetal2022,GrishinPerets2022,Stegmann2024}. While triples with an outer semi-major axis of $a_{\rm out} \lesssim \mathcal{O}\left(10^4\right)$ AU are not expected to be disrupted by the Galactic tide \citep{Jiang2010,ElBadry2018}; even for outer semi-major axes of order $10^{3\textrm{--}4}$AU the tertiary may be frequently scattered to a highly eccentric orbit about the inner binary \citep{Stegmann2024}. Thus, the hierarchical triple may enter such a state that the hierarchy of orbital periods is still preserved, but the outer pericentre---while still larger than the inner semi-major axis---is much smaller than the outer semi-major axis. We call such triples \emph{quasi-hierarchical}. 

These triples cannot be modelled as regular hierarchical systems, because the time the tertiary spends near pericentre is comparable to (a few times) the inner orbital period. While fully hierarchical systems can be studied perturbatively, and fully democratic ones can be described probabilistically, quasi-hierarchical triples, which inhabit the intermediate range, do not succumb to either treatment: the separation of time-scales is not strong enough for standard triple perturbation theory, but there is still too much structure (not enough phase-space mixing) for a purely statistical-mechanical theory. 
In terms of eccentricity, we will show below in appendix \ref{appendix: random walk stuff} that if the outer orbit's eccentricity is $e_{\rm out} \gtrsim 1-\sqrt{a_{\rm in}/a_{\rm out}}$, then the triple becomes quasi-hierarchical. If $e_{\rm out} \gtrsim 1-a_{\rm in}/a_{\rm out}$ the triple becomes unstable \citep{MardlingAarseth2001}. 

The quasi-hierarchical regime is distinct from other deviations from secularity, where, e.g., double averaging fails because the time-scale separation is not wide enough \citep{Luoetal2016,Grishin2018,Mangipudietal2022}. This can be corrected, yielding the Brown Hamiltonian \citep[e.g.][]{Luoetal2016,Will2021,Tremaine2023,Grishin2024,KleinKatz2024,LeiGrishin2025a,LeiGrishin2025b}, but these studies still took $e_{\rm out}$ to be moderate, and kept higher-order terms in $\alpha \equiv a_{\rm in}/a_{\rm out}$. The quasi-hierarchical regime here is distinct: the time-scale hierarchy is very wide, but on the other hand $e_{\rm out}$ is very high, too. 
As perturbers excite $e_{\rm out}$, one thus has three possibilities for the subsequent evolution, summarised in Table~\ref{tab:regimes}. 

\begin{table*}
    \centering
        \begin{tabular}{|l|c|l|}
        \hline
            Main dynamical evolution & range of $e_{\rm out}$ & probability for thermal distribution (with $a_{\rm out} = 100a_{\rm in}$) \\
            \hline
            Secular & $1-e_{\rm out} \geq \sqrt{a_{\rm in}/a_{\rm out}}$ & $\mathcal{O}(1)$~~~~~~~~~~~~~~~~~~~~~($81\%$) \\
            Quasi-hierarchical (this work) & $\sqrt{a_{\rm in}/a_{\rm out}} \geq 1- e_{\rm out} \geq a_{\rm in}/a_{\rm out}$ & $\mathcal{O}(\sqrt{a_{\rm in}/a_{\rm out}})$~~~~~~~~\,($17\%$) \\
            Unstable (strong triple scattering) & $1 - e_{\rm out} \leq a_{\rm in}/a_{\rm out}$ & $\mathcal{O}(a_{\rm in}/a_{\rm out})$~~~~~~~~~~~\,($2\%$) \\
            \hline
        \end{tabular}
    \caption{Three regimes for dynamical triple evolution. The right column displays the relative probability of being in each one, for a thermal distribution of $e_{\rm out}$ ($p(e_{\rm out}) = 2e_{\rm out}$), and $a_{\rm in}/a_{\rm out} = 0.01$.}
    \label{tab:regimes}
\end{table*}

The aim of this paper is to introduce a simple model for the evolution of such a triple, that tracks the inner binary's orbital elements over time. This model will allow us to gauge the time-scale for the evolution of the inner binary's eccentricity (\emph{inter alia}), and hence how fast it reaches values where other effects, such as stellar tidal interactions, stellar collisions, or gravitational-wave emission, become important. 

Indeed, since the first direct detection of gravitational waves from a binary-black-hole coalescence by \cite{LIGOVirgo2016}, many channels have been proposed for the origin of the binaries seen by the LIGO--Virgo--KAGRA (LVK) collaboration, with the goal of explaining how pairs of compact objects can be brought to tight separations, that allow them to merge within a time shorter than the age of the Universe. These channels include globular or nuclear clusters 
\citep[e.g.,][]{ZwartMcMillan2000,OLearyetal2009,Tanikawa2013,AntoniniRasio2016,Rodriguezetal2015,Rodriguezetal2016,Samsing2018,SamsingDorazio2018,Antoninetal2025}, active galactic nuclei (AGN) \citep[e.g.,][]{Bartosetal2017,Stoneetal2017,Tranietal2019,Tranietal2021,Tranietal2024,McKernan2020,Tagawaetal2020,Samsingetal2022,Samsingetal2024,RoznerPerets2022,Grishinetal2024,FabjSamsing2024,Whiteheadetal2024,Gilbaumetal2025,Rowanetal2025}, an isolated channel where the tight binaries are formed via binary stellar evolution \citep[e.g.,][]{Belczynski2002,Dominiketal2012,Belczynskietal2016,Iorio2023}, and a ``triple" channel 
\citep[e.g.,][]{Wen2003,Antognini2014,Antonini2014,Antonini2016,Antonini2017,Rodriguez2018,FragioneLoeb,Fragioneetal2019,Bartos2023,Vigna2025,Stegmann2025}, where the black-hole binary is the inner component of a hierarchical triple system, and the tertiary acts as a reservoir with which the inner binary may exchange angular momentum. This sets the triple channel apart from the others, where the binary shrinks by giving its energy to the reservoir (the cluster, the AGN or the envelope of its companion); in the triple channel, energy is not exchanged with the outer orbit---only angular momentum (in other dynamical channels angular momentum is of course exchanged, too, but energy also is).\footnote{Although an energy exchange may occur in gaseous environment \citep{Suetal2025}.} Once the inner binary becomes eccentric enough, it releases its energy directly into gravitational waves. 

The paper is organised as follows: we start by describing an analytical map that models the evolution of the orbital elements of a quasi-hierarchical triple in \S \ref{sec:quasi-hierarchical triples}, which we then show leads to a somewhat different maximum secular eccentricity than ZLK oscillations (\S \ref{sec:secular}), and we comment on its consequences for gravitational-wave coalescences. Then, we consider the long-term evolution of triples under the map, approximating as a random walk in \S \ref{sec:external}, and describe some of its consequences. We conclude by summarising our findings in \S \ref{sec:discussion}.

\section{Evolution of Quasi-Hierarchical Triples}
\label{sec:quasi-hierarchical triples}
As remarked in the introduction (\S \ref{sec:introduction}), a quasi-hierarchical triple is a triple system where the semi-major axis $a_{\rm in}$ of the inner binary (composed of masses $m_1$ and $m_2$) is much smaller than that of the outer binary (comprising the inner binary's centre of mass and the mass $m_3$), denoted by $a_{\rm out}$; however, the eccentricity of the tertiary's orbit about the inner binary centre-of-mass is so large, that its periapsis, $r_{\rm p} \equiv r_{\rm p, out}$ is not much smaller than $a_{\rm in}$. Thus, we have 
\begin{equation}\label{eqn:fundamental assumption}
	1 \lesssim \frac{r_{\rm p}}{a_{\rm in}} \ll \frac{a_{\rm out}}{a_{\rm in}}.
\end{equation}

In this case, the inner binary's evolution is dominated by the `impulsive kicks' it receives during each pericentre passage of the outer object \citep{Antoninietal2010}. Each of these close encounters is similar to an interaction of a binary with an unbound perturber on a parabolic orbit, whose distance of closest approach is $r_{\rm p}$; such close encounters have already been studied in the literature by \cite{HeggieRasio1996,HamersSamsing2019a,HamersSamsing2019b}. Here, the binary evolves under a sequence of such kicks, which are correlated with each other, in the sense that they are all determined by the same outer orbit. 

\begin{figure}
    \centering
    \includegraphics[width=0.48\linewidth]{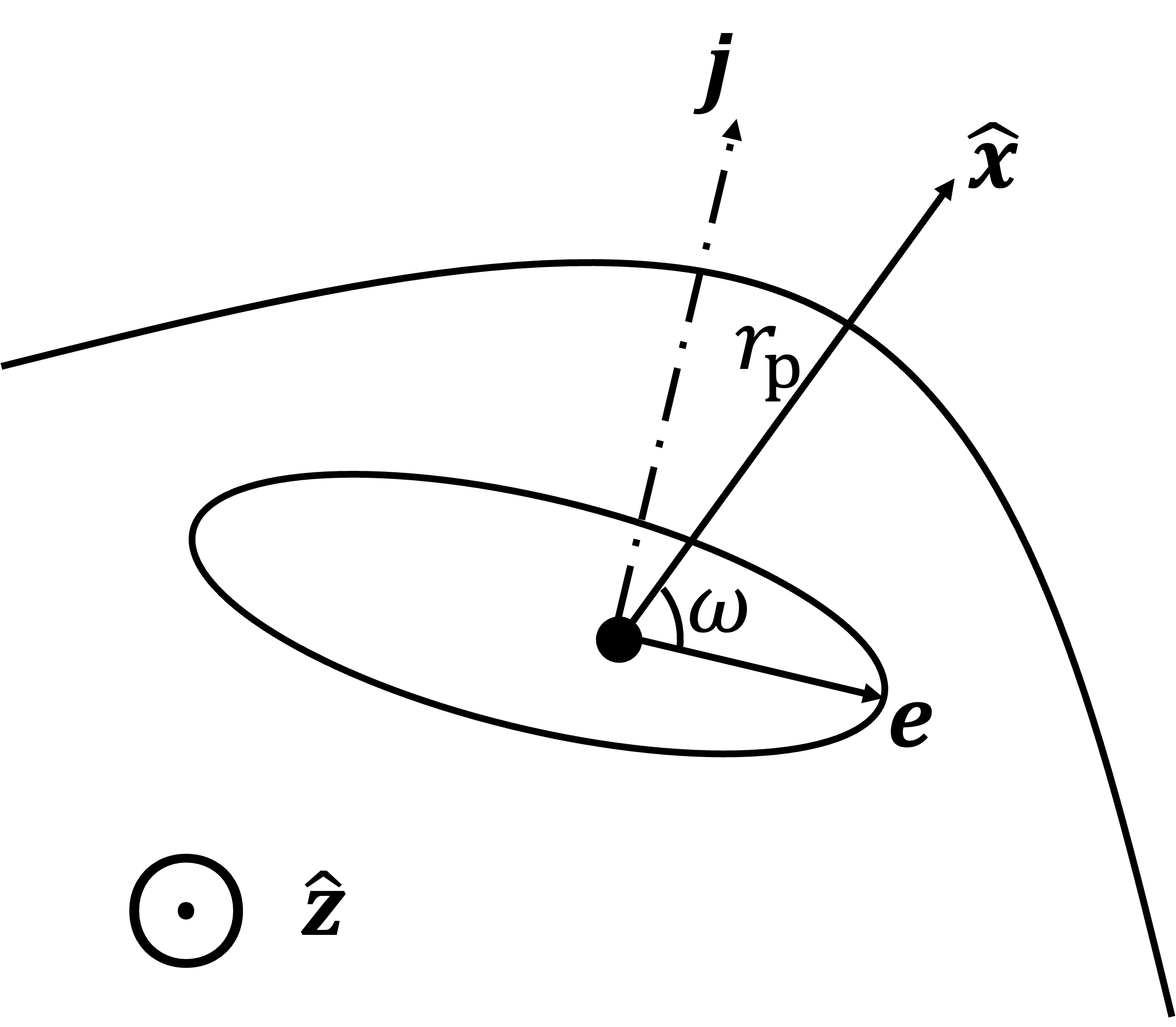}
    \caption{A depiction of the orientations of the three bodies (see text and equations \eqref{eqn:vectors}). The $\xhat$-axis points from the binary's centre-of-mass to the outer pericentre, and the $\zhat$-axis points out of the page.}
    \label{fig:angles}
\end{figure}

The system is still hierarchical, so one may average the interaction between the binary and the outer body over the period of the inner binary---but not over the period of the outer. This implies that the energy of the inner binary is conserved, but its angular momentum is not. The latter is characterised by the Laplace--Runge--Lenz vector 
\begin{equation}\label{eqn:vectors}
	\mathbf{e} \equiv e \left( \begin{array}{c}
		\cos \Omega \cos \omega - \sin \Omega \sin \omega \cos i \\
		\sin \Omega \cos \omega + \cos \Omega \sin \omega \cos i \\ 
		\sin i\sin \omega 
	\end{array}\right),
\end{equation}
and the dimension-less angular momentum 
\begin{equation}\label{eqn: j definition}
	\mathbf{j} \equiv \sqrt{1-e^2}\left( \begin{array}{c}
		\sin \Omega \sin i \\
		-\cos \Omega \sin i \\ 
		\cos i 
	\end{array}\right),
\end{equation}
where $i\equiv \arccos\left(\mathbf{j}\cdot\zhat\right)$ and have adopted the angle conventions of \cite{HamersSamsing2019a}, where the outer angular momentum is always $\mathbf{J}_{\rm out} \parallel \zhat$. We fix the outer orbit's orientation as depicted in figure \ref{fig:angles}.
With the semi-major axis $a_{\rm in}$ and the masses being fixed, $\mathbf{e}$ and $\mathbf{j}$ determine the inner binary's orbit completely. 
After each close pericentre passage of the outer body, $\mathbf{e}$ changes by an amount $\Delta \mathbf{e}$, and $\mathbf{j}$ changes by $\Delta \mathbf{j}$. The magnitude of these changes is determined by the hierarchy parameter \citep{HamersSamsing2019a} 
\begin{equation}\label{eqn: hierarchy parameter definition}
	\eps \equiv \sqrt{\frac{m_3^2}{m_bM}\left(\frac{a_{\rm in}}{r_{\rm p}}\right)^3},
\end{equation}
where $m_{\rm b}\equiv m_1+m_2$ is the inner binary's mass and $M \equiv m_{\rm b} + m_3$ is the total mass. We denote the inner (outer) reduced mass by $\mu_{\rm in}$ ($\mu_{\rm out}$).

Furthermore, because $r_{\rm p} \ll a_{\rm out}$, we may approximate the outer eccentricity as $e_{\rm out} \approx 1$; that is, we will treat each of the close pericentre passages as a parabolic encounter. Under this assumption, 
\begin{equation}\label{eqn: changes per encounter}
	\begin{aligned}
		\Delta \mathbf{e} & = \frac{3\pi}{2}\eps \left(\begin{array}{c}
			-e_zj_y - e_yj_z \\
			e_z j_x + e_x j_z \\
			2e_y j_x - 2e_x j_y
		\end{array}\right) + \eps^2 \mathbf{g}_e, \\  
		\Delta \mathbf{j} & = \frac{3\pi}{2}\eps \left(\begin{array}{c}
			-j_yj_z - 5e_ye_z \\
			5e_x e_z - j_x j_z \\
			0
		\end{array}\right) + \eps^2 \mathbf{g}_J,
	\end{aligned}
\end{equation}
where the second-order corrections $\mathbf{g}_e$ and $\mathbf{g}_J$ are given by equation~(27) of \cite{HamersSamsing2019a}, at quadrupole order. 
In particular, the change in eccentricity, $\Delta e = (\mathbf{e} \cdot \Delta \mathbf{e})/e$ reads
\begin{equation}\label{eqn: eccentricity change second order}
	\begin{aligned}
		\Delta e & = \frac{15\pi}{4}\eps e\sqrt{1-e^2}\sin^2 i \sin 2\omega + \frac{9\pi^2}{512}\eps^2 e \left(124-299 e^2\right)\\ &
		+ \frac{3\pi}{512}\eps^2 e  \bigg\{100 \left(1-e^2\right) \sin 2 \omega  \big[(5 \cos i+3 \cos 3 i) \cos 2 \Omega \\ & \quad 
		+6 \sin i \sin 2 i\big] +4 \cos 2 i \left[200 \left(1-e^2\right) \cos 2 \omega  \sin 2 \Omega \right. \\ & \quad
		\left. + 3 \pi  \left(81 e^2-56\right)\right]+3 \pi  \left[200 e^2 \sin ^4i \cos 4 \omega \right.\\ & \quad
		\left.+8 \left(16 e^2+9\right) \sin ^2 2i \cos 2 \omega +\left(39 e^2+36\right) \cos 4 i \right]\bigg\}.
	\end{aligned}
\end{equation}

The evolution of the system is therefore approximated by a sequence of $N$ parabolic encounters, each of which change the inner binary orbit as 
\begin{equation}\label{eqn: evolution}
	\begin{aligned}
		a_{n+1} & = a_n, \\ 
		\mathbf{e}_{n+1} & = \mathbf{e}_n + \Delta\mathbf{e}(a_n,e_n,i_n,\omega_n,\Omega_n) ,\\ 
		\mathbf{j}_{n+1} & = \mathbf{j}_n + \Delta\mathbf{j}(a_n,e_n,i_n,\omega_n,\Omega_n).
	\end{aligned}
\end{equation}
We account for angular-momentum conservation explicitly, and for the possibility of orbit flips as explained in appendix \ref{appendix: details quasi hierarchical}. In that appendix, we also describe our implementation of the method of \cite{Macguireetal2026} to account for the phase-space constraints $\mathbf{e}\cdot \mathbf{j} = 0$ and $e^2 + j^2 = 1$.

This approach is empirically found to be valid for $r_{\rm p} \gtrsim 3a_{\rm in}$ \citep{Samsingetal2019} in the case of a single pericentre passage; below this ratio, even the single-averaging over the inner binary is inadequate, so energy exchanges need to be accounted for, and incorporated into the kicks \citep{MushkinKatz2020}. We do not consider such cases here, and restrict ourselves to the case where averaging over the inner binary's orbit---and hence $a_{n+1} = a_n$---is still acceptable. 
Appendix \ref{appendix: details quasi hierarchical} provides more details of the behaviour of $\Delta e$ engendered by this map, as a function of the various angles involved.

\section{Secular Oscillations}
\label{sec:secular}
The map \eqref{eqn: evolution} leads to secular oscillations of $e_n$. Figure shows a comparison of the outcome of applying it to a numerical integration of the equations of motion with direct three-body simulations. 

\begin{figure*}
	\centering
	\includegraphics[width=0.49\textwidth]{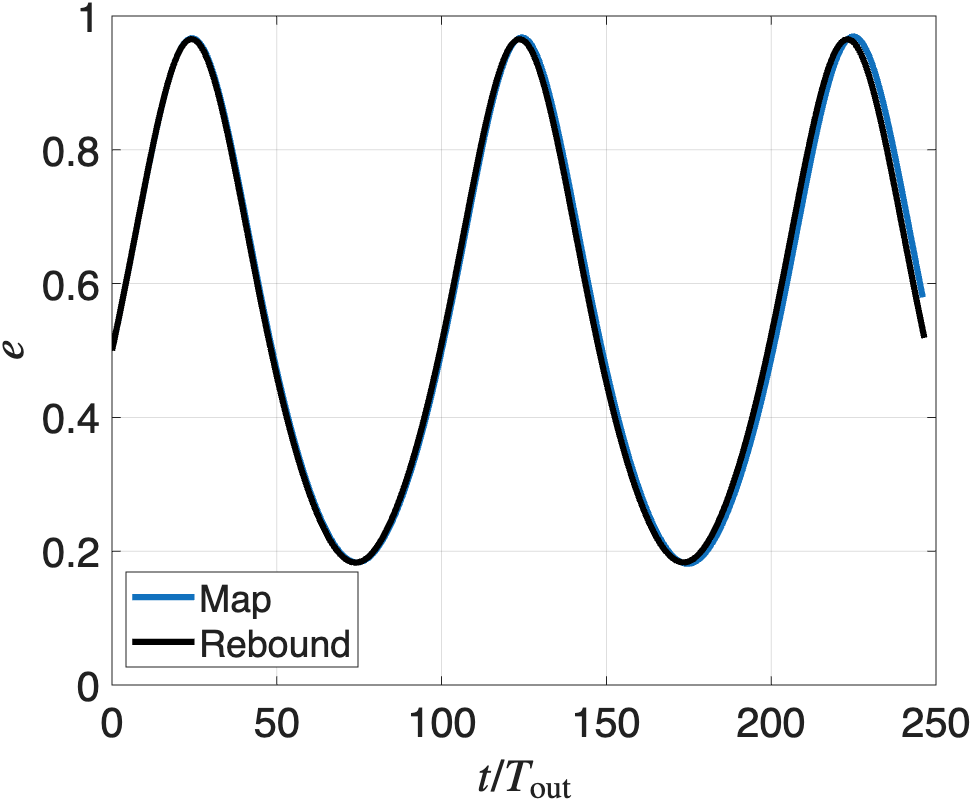}
	\includegraphics[width=0.49\textwidth]{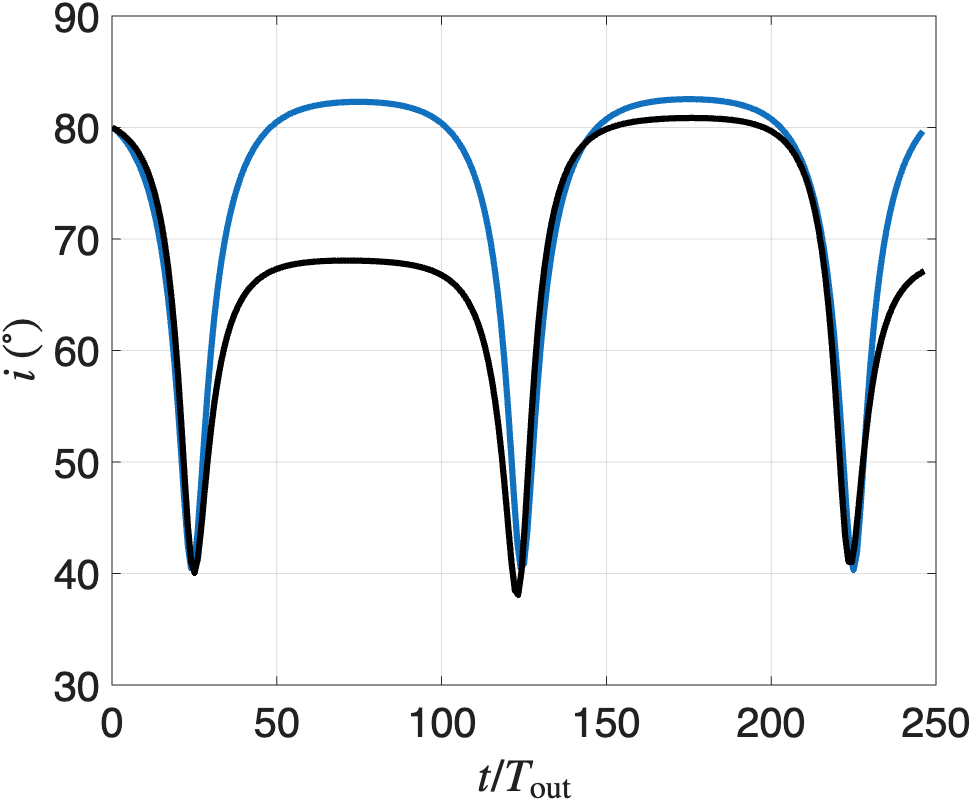}
	\includegraphics[width=0.49\textwidth]{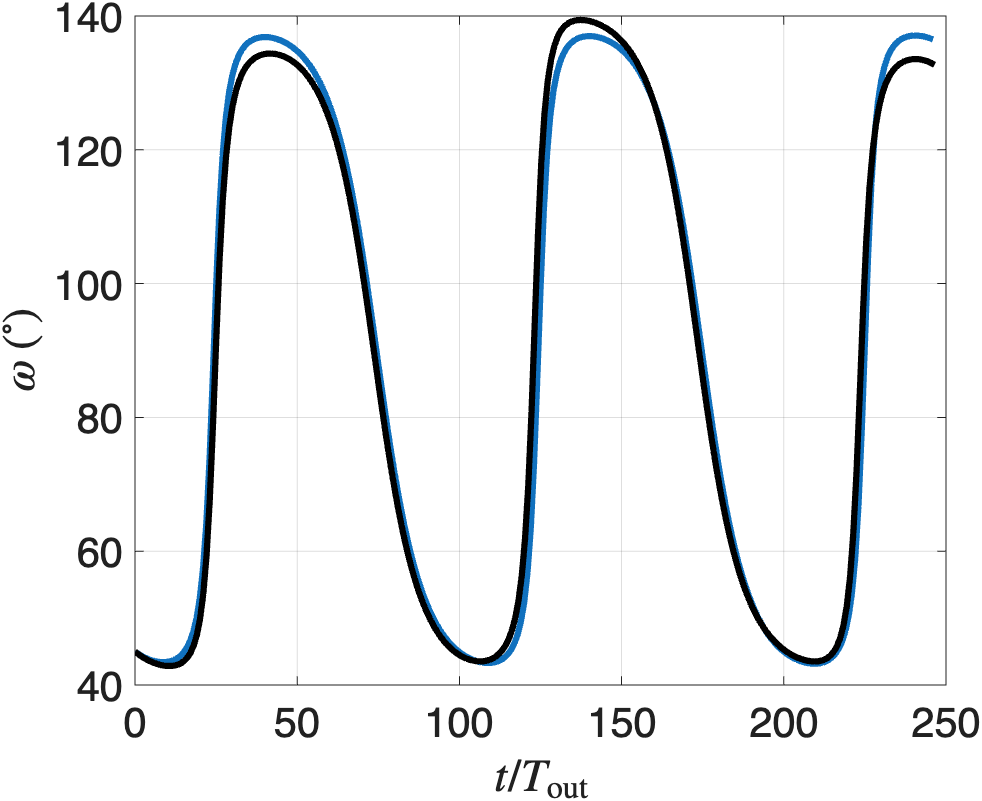}
	\includegraphics[width=0.49\textwidth]{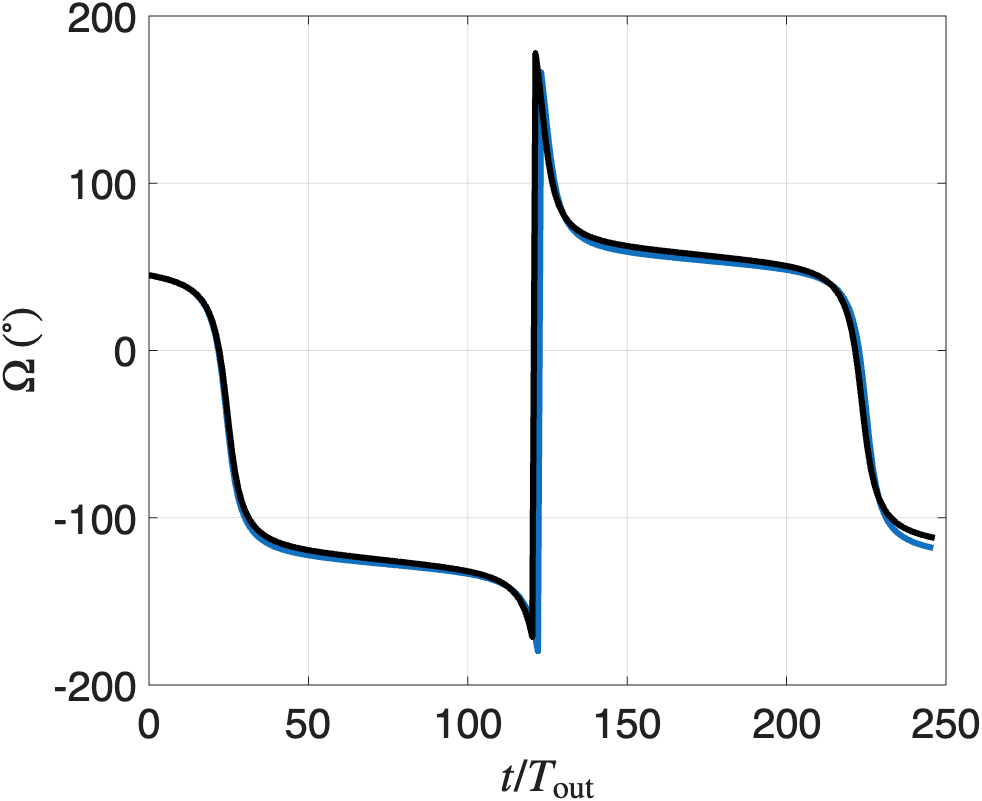}
	\caption{A comparison between the analytical map (blue) $\Delta \mathbf{e}$, $\Delta \mathbf{j}$ given in equations (\ref{eqn: changes per encounter}--\ref{eqn: evolution}), and the results of a direct three-body simulation (in black, dash-dotted); see text for details. We see that the simulations matches the analytical formula well. We ascribe the disagreement in inclination to a different way of defining it in {\tt rebound}.}
	\label{fig:rebound}
\end{figure*}

We compare the analytical prescription to three-body simulations run with the \verb"rebound" code \citep{ReinLiu2012,ReinTamayo2015}, using the \verb"ias15" integrator, with the following configuration: $m_1=m_2=m_3=M_\odot$, $a_{\rm in} = 1$AU, $a_{\rm out} = 1000a_{\rm in}$, $e_{\rm out} = 0.99$, $i_0 = \pi/2$, $\omega_0 = \Omega_0 = \pi/4$. We run the code for $250$ outer orbits, and compare the evolution of orbital parameters of the inner orbit with those predicted by equations~\eqref{eqn: evolution}. The comparison, showing good agreement over many outer orbits---for both the magnitude of the changes and the time-scale---is displayed in figure \ref{fig:rebound}.\footnote{We believe the discrepancy in the inclination to arise from an inconsistency between our definition for it, and the one used in {\tt rebound}. We did attempt to understand if this is indeed a geometric problem, but were unable to find a corrected definition that resolved it. We still suspect it to be the case, though, because all other angles agree very well, and because the inclination does return to agreement with the map after about $120 T_{\rm out}$.} While we do not plot $a_{\rm in}$ over multiple outer-pericentre passages, the \texttt{rebound} simulations do conserve it as expected. We see that the map traces the evolution of the system faithfully for hundreds of outer orbits---a few secular cycles. Then, the two descriptions start to deviate. The discrepancy arises from the impact of $\mathcal{O}(\eps^3)$ and $\mathcal{O}(1-e_{\rm out})$ terms that we have neglected, over long times.

The secular time-scale can be immediately gleaned from equation \eqref{eqn: eccentricity change second order}:
\begin{equation}
\begin{aligned}
  & \frac{T_{\rm out}}{\tau_{\rm sec}} \sim \frac{15\pi \eps e \sqrt{1-e^2}\sin^2i}{4} + \frac{9\pi^2}{512}\eps^2 e \\ & 
  \times \left[4(81e^2-56)\cos 2i + (39e^2+36)\cos 4i - 299e^2 + 124\right]\,,
\end{aligned}
\end{equation}
where the second line matters mostly for $i\sim 0^\circ , 180^\circ$, for it is otherwise smaller than the first. To leading order, 
\begin{equation}
    \tau_{\rm sec} \sim T_{\rm out} \frac{\sqrt{m_bM}}{m_3 e\sqrt{1-e^2}\sin^2i}\left(\frac{r_{\rm p}}{a_{\rm in}}\right)^{3/2}\,;
\end{equation}
this is the eccentric ZLK time-scale, divided by the denominator $e\sqrt{1-e^2}\sin^2i$. 

\begin{figure}
    \centering
    \includegraphics[width=0.48\textwidth]{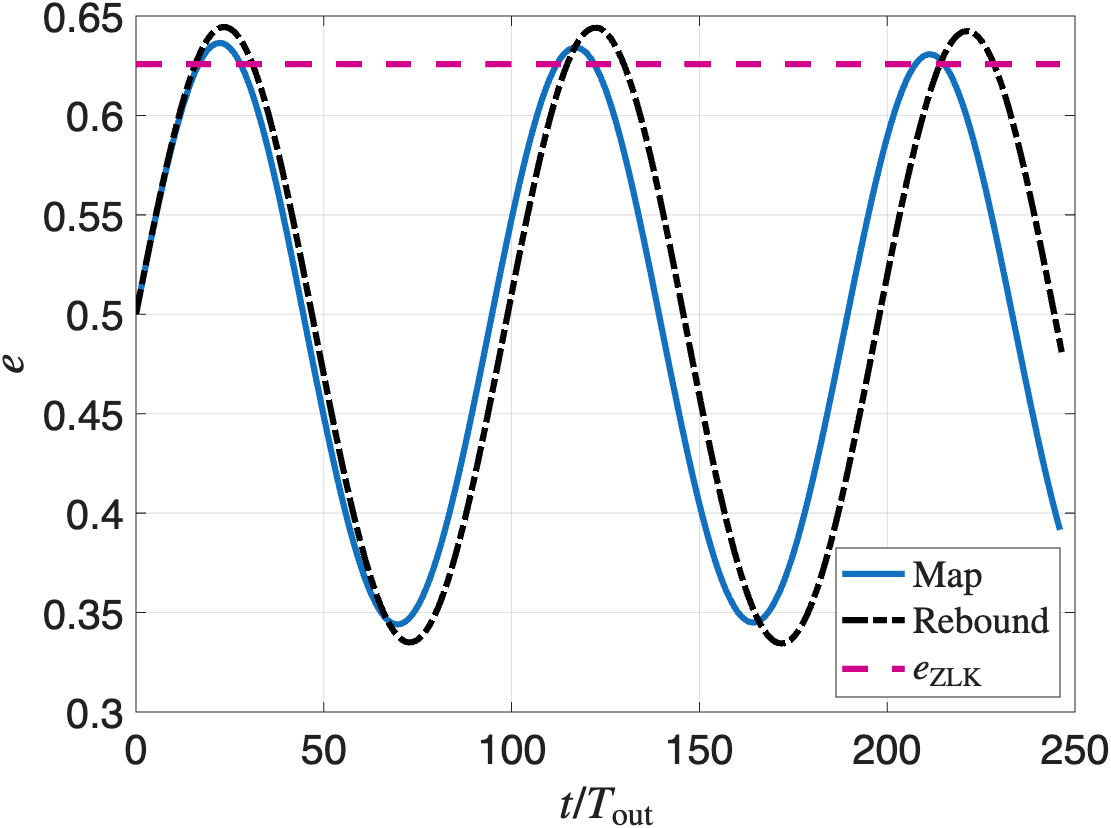}
    \includegraphics[width=0.48\textwidth]{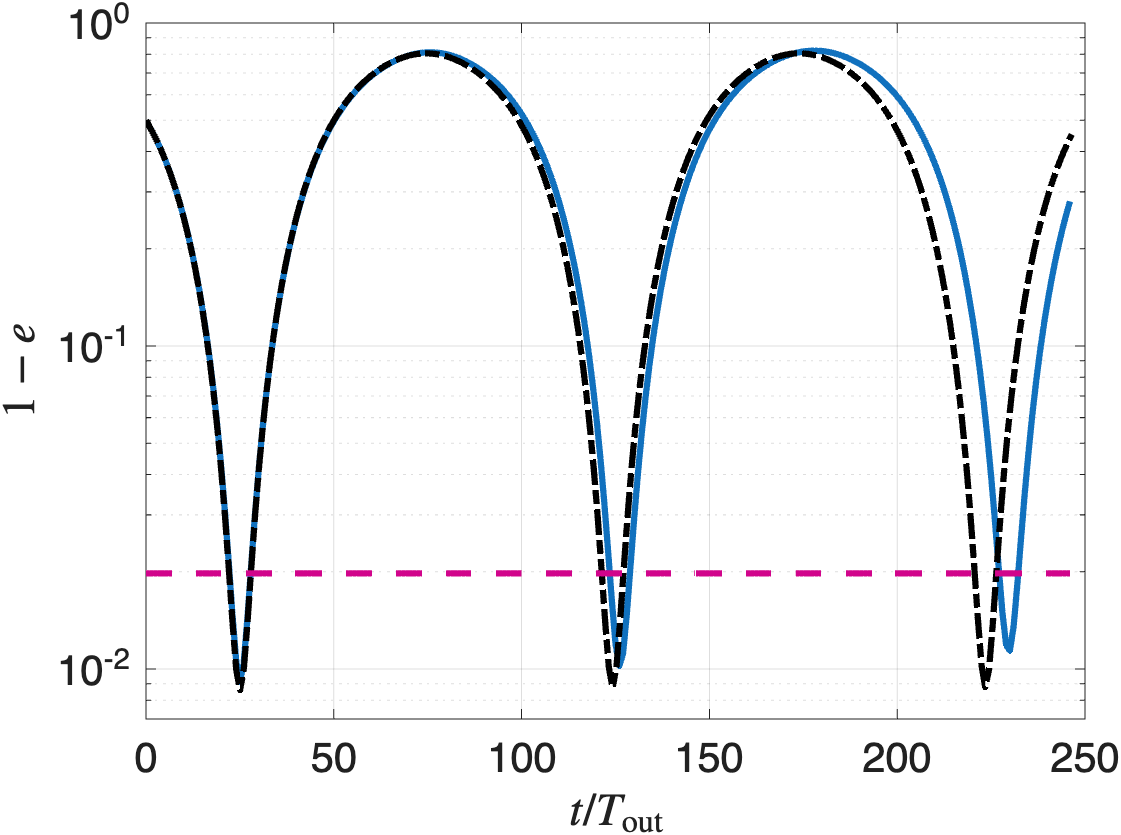}
    \caption{The evolution of a quasi-hierarchical orbit over a few secular times, from the map \eqref{eqn: evolution} (blue) and {\tt rebound} (black), for $m_1 = m_2 = m_3 = M_{\odot}$, $\omega_0 = \Omega_0 = \pi/4$, $r = 10a_{\rm in}$, $a_{\rm out} = 1000a_{\rm in}$; \emph{top}: $i_0 = 7\pi/9$, \emph{bottom}: $i_0 = 5\pi/9$. The purple curve shows is the maximum eccentricity according to pure quadrupole ZLK evolution.}
    \label{fig:e_t comparison exceeding emax}
\end{figure}

We can compute the maximum eccentricity $e_{\max}$ that the triple can achieve over a secular cycle, by a repeated iteration of equation \eqref{eqn: evolution}. This can be compared with predictions from the quadrupole ZLK mechanism \citep[e.g.][]{Naoz2016,HamiltonRafikov2019}. As seen from an example in figure \ref{fig:e_t comparison exceeding emax}, quasi-hierarchical triples sometimes exceed the ZLK maximum eccentricity. 

\begin{figure*}
    \centering
    \includegraphics[width=0.49\textwidth]{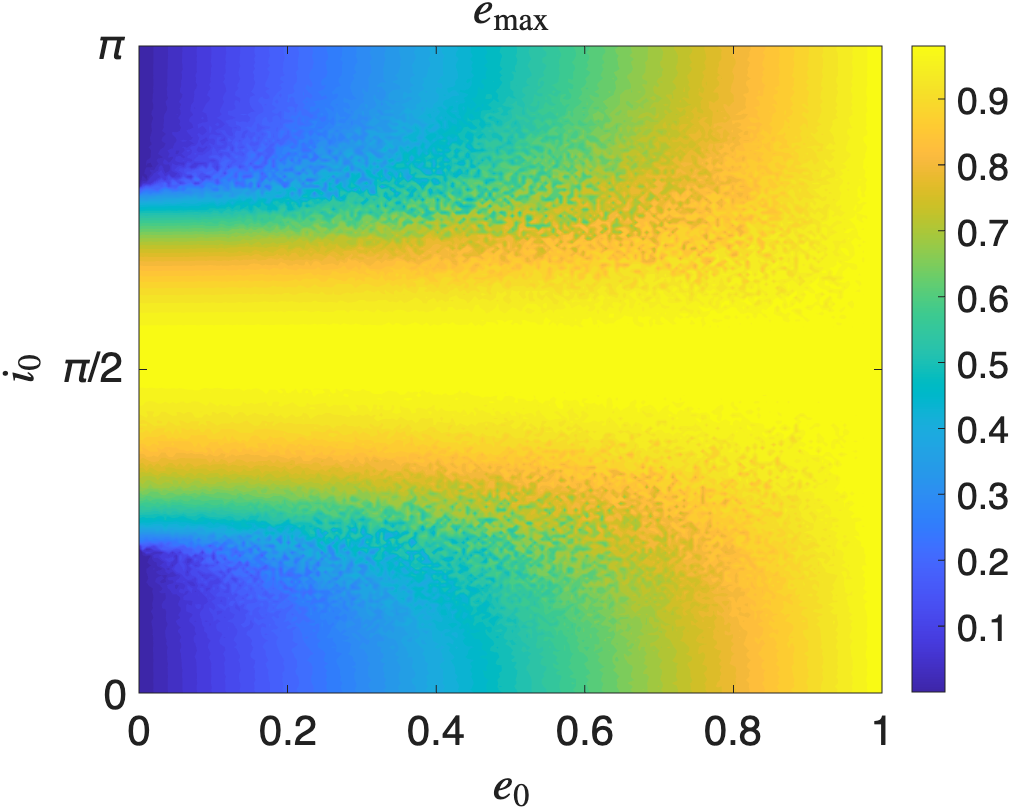}
    \includegraphics[width=0.495\textwidth]{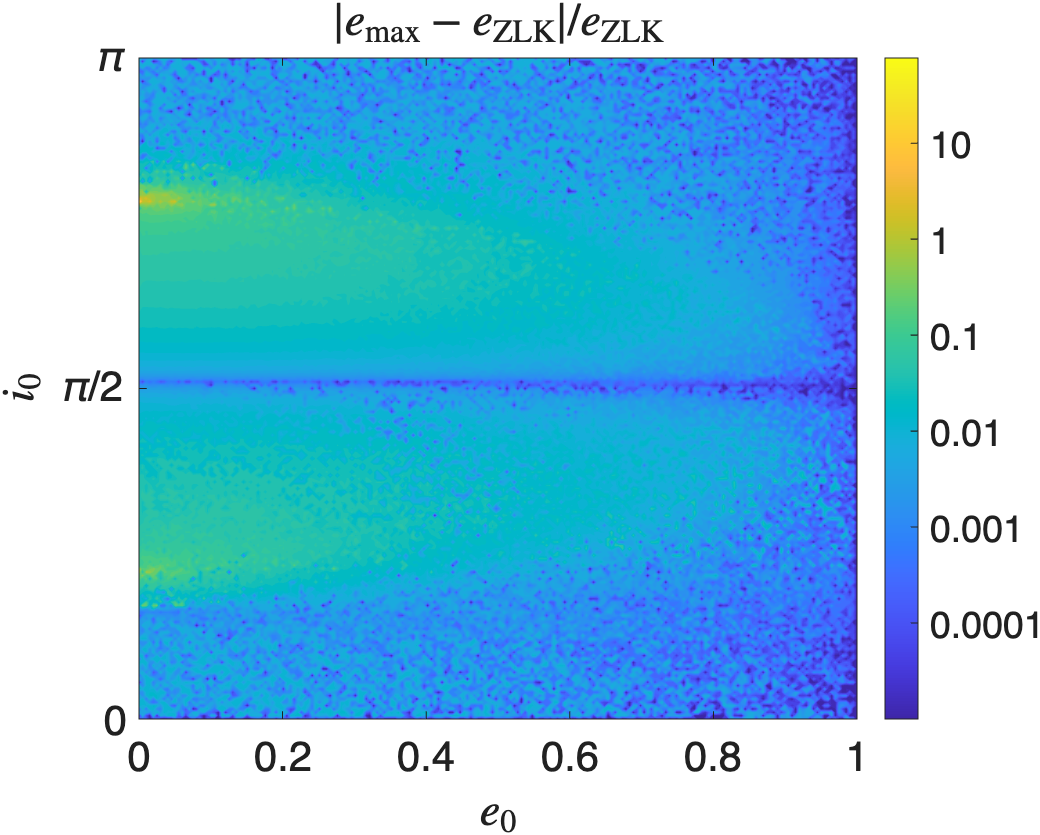}
    \caption{\emph{Left}: a diagram of the maximum eccentricity achieved by a quasi-hierarchical triple over an $\mathcal{O}(\tau_{\rm sec})$ time-scale, found as the first local eccentricity maximum, for equal masses. \emph{Right}: a comparison with the maximum eccentricity according to pure (quadrupole) ZLK evolution; the colour-scheme is logarithmic. See text for details.}
    \label{fig:emax histogram}
\end{figure*}

\begin{figure}
    \centering
    \includegraphics[width=0.48\textwidth]{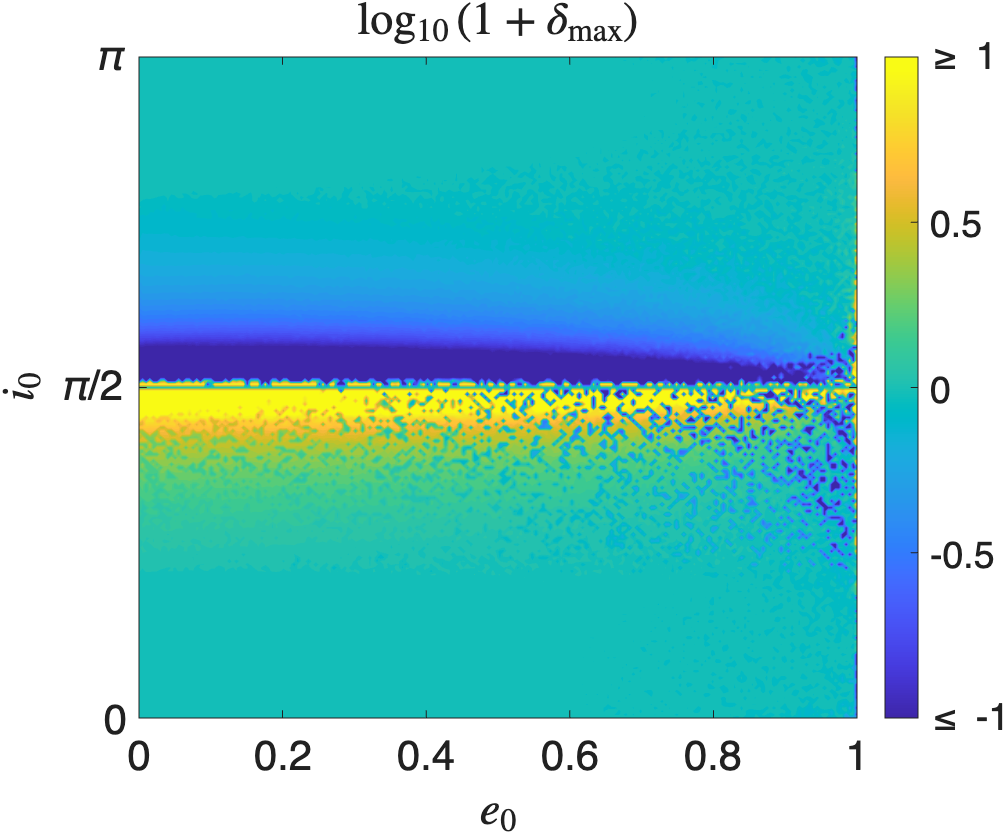}
    \caption{The change in the time to coalescence from standard ZLK evolution, from equation \eqref{eqn:delta gw}. The change is stronger around $i_0 = \pi/2$, where both $e_{\max}$ and $e_{\rm ZLK}$ are very high.}
    \label{fig:delta gw}
\end{figure}

In figure \ref{fig:emax histogram} we explore how the value of $e_{\max}$ depends on the initial eccentricity and inclination. As can be seen from the right panel, the deviation is sometimes significant. 
The difference in maximum eccentricity leads to a difference in the time to coalescence due to gravitational-wave emission. For an orbit with eccentricity $e$, it is given by 
\begin{equation}
  \tau_{\rm gw} = \tau_{\rm c}\left(1-e^2\right)^{7/2}\,,
\end{equation}
where $\tau_{\rm c}$ is the time-to-coalescence of a circular binary, given by \citep{Maggiore}
\begin{equation}\label{eqn:time-to-coalescence circular}
	\tau_{\rm c} = \frac{5}{256}\frac{c^5 a_{\rm in}^4}{Gm_{\rm c}^{5/3}m_{\rm b}^{4/3}}\,.
\end{equation}
\cite{LiuLai2017,LiuLai2018} argued that given the strong dependence of $\tau_{\rm gw}$ on the eccentricity, one could approximate the entire energy loss as occurring near pericentre. As $e$ fluctuates, the only time any significant GW emission occurs, is when $e \sim e_{\max}$. Thus, according to \cite{LiuLai2017,LiuLai2018}, the time a binary, undergoing secular oscillations, takes to coalesce is well approximated by
\begin{equation}
  t_{\rm coal} \approx \tau_{\rm c}\left(1-e_{\max}^2\right)^{7/2} \frac{T(e \sim e_{\max})}{\tau_{\rm sec}(\textrm{low } e)}\,,
\end{equation} 
where $T(e \sim e_{\max})$ is the time the secular oscillations spend with $e$ near $e_{\max}$. For ZLK evolution this is $\tau_{\rm sec}(\textrm{low } e) \sqrt{1-e_{\max}^2}$, and here, we estimate it as follows: for strong GW emission we need $e_{\max} - e \ll 1-e_{\max}$, whence equation \eqref{eqn: eccentricity change second order} implies that here
\begin{equation}
  T(e \sim e_{\max}) \sim \tau_{\rm sec}(\textrm{low } e) \frac{\sqrt{1-e_{\max}^2}}{e_{\max}}\sin^2 i(e_{\max})\,,
\end{equation} 
retaining the same $\sqrt{1-e_{\max}^2}$ dependence as the ZLK case. Therefore, neglecting the inclination dependence, the \cite{LiuLai2017,LiuLai2018} relation $t_{\rm coal} = \tau_{\rm c}\left(1-e_{\max}^2\right)^3$ obtains here, too; quantitatively, it is modified by
\begin{equation}\label{eqn:delta gw}
  \frac{t_{\rm coal, ZLK}}{t_{\rm coal}} \approx 1+\delta_{\max} \equiv \left(\frac{1-e_{\rm ZLK}^2}{1-e_{\max}^2}\right)^3. 
\end{equation}
This is plotted in figure \ref{fig:delta gw}. The deviation is largest around $i_0 \sim 90^\circ$, because that's where $t_{\rm coal}$ is shortest, and can be significant. 

\section{External Effects and Possible Eccentricity Drift}
\label{sec:external}
The angular-momentum changes accumulate over long times, leading to a quasi-random behaviour of the map \eqref{eqn: evolution}, as can be shown in figure \ref{fig:example evolution}. This figure was created using random initial conditions, with $r_{\rm p} = 10a_{\rm in}$, and equal binary masses. We terminate the evolution once $e_n$ reaches $1$. We display five example trajectories. 

\subsection{Pure evolution under the map}
\begin{figure*}
	\centering
	\includegraphics[width=0.49\linewidth]{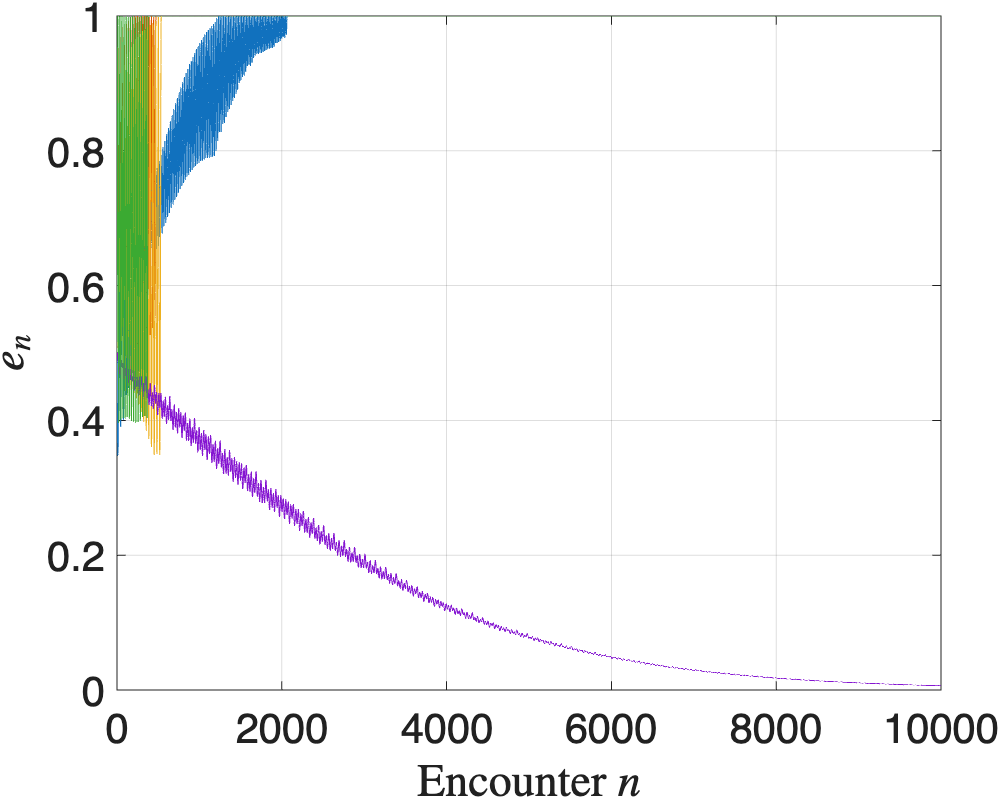}
	\includegraphics[width=0.49\linewidth]{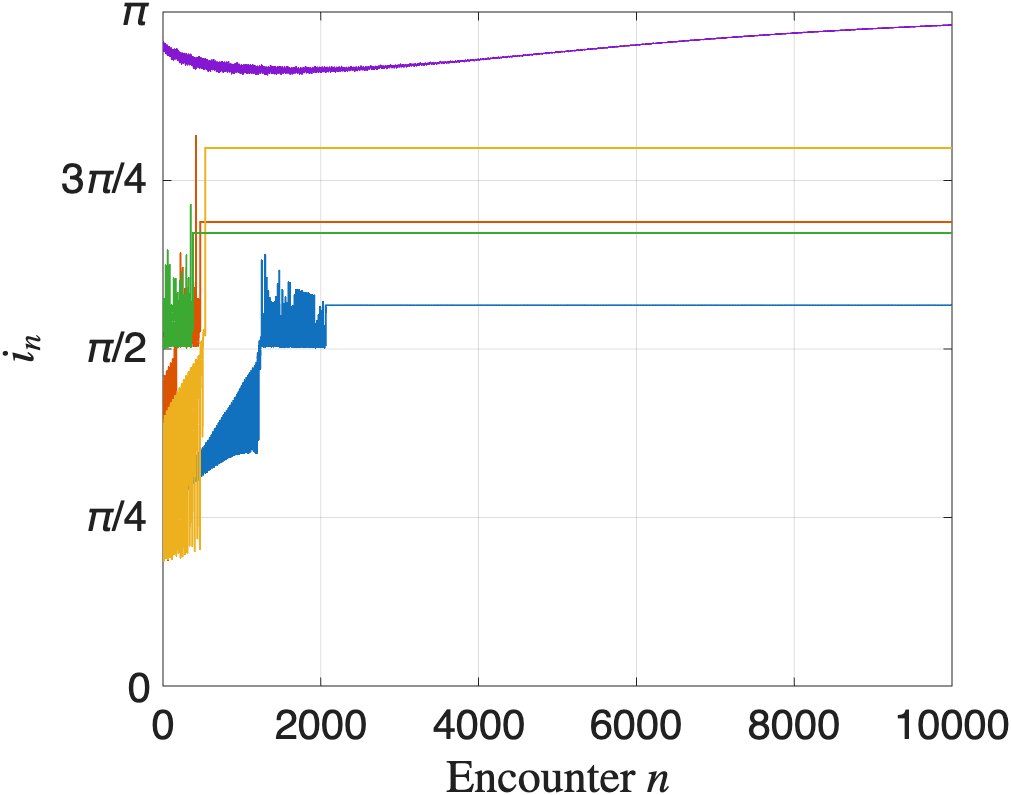}
    \includegraphics[width=0.49\linewidth]{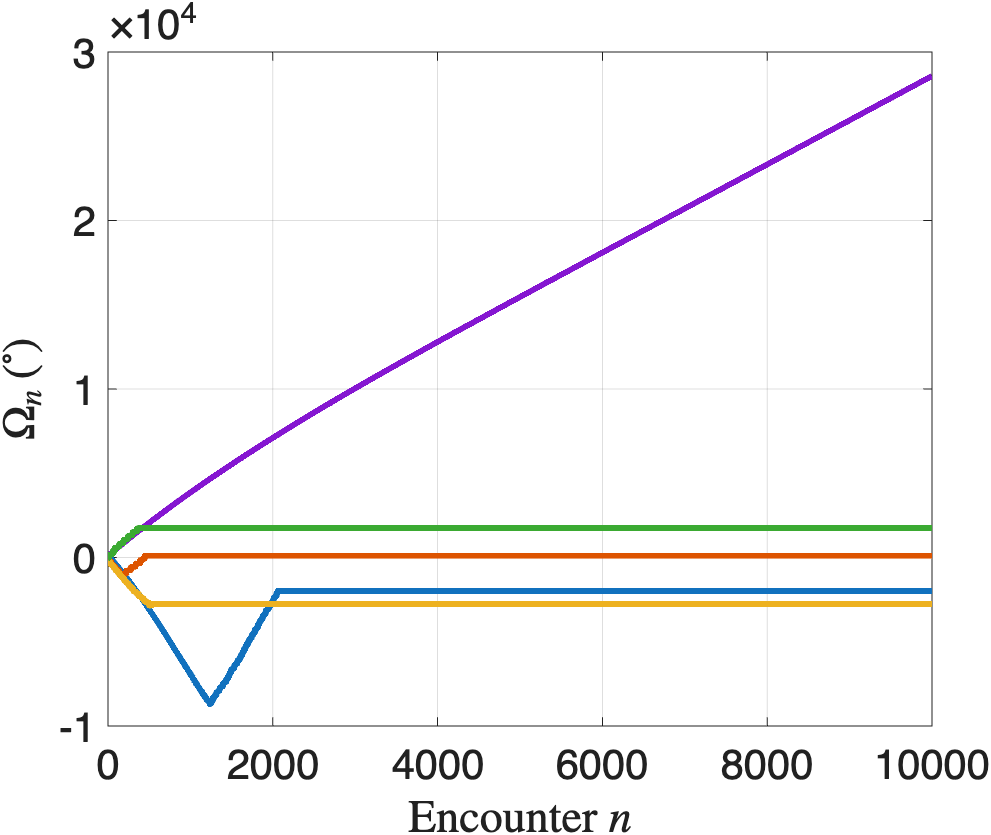}
    \includegraphics[width=0.5\linewidth]{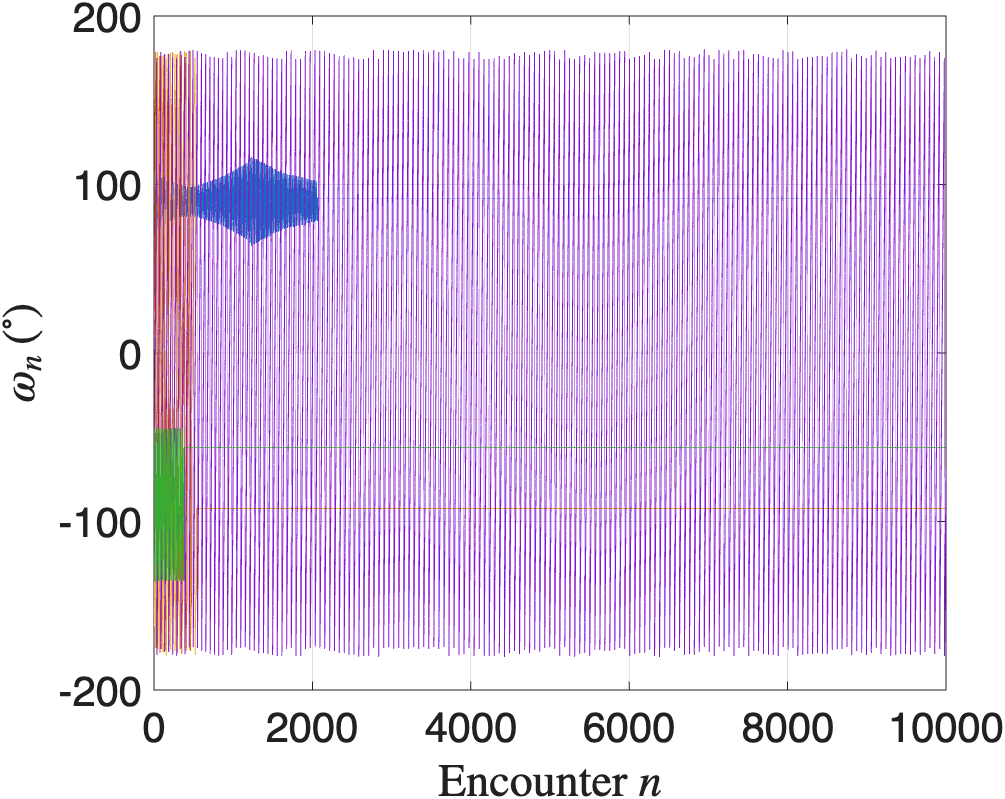}
	\caption{The evolution of randomly chosen initial conditions, for an equal-mass triple, for $10000$ outer orbits, and $r_{\rm p} = 10a_{\rm in}$, as determined by equations~\eqref{eqn: changes per encounter}. The system is frozen when $e_n$ reaches $0.9999$. See text and appendix \ref{appendix: details quasi hierarchical} for details. The colours correspond to five random realisations of the initial conditions, and are consistent across the four panels.}
	\label{fig:example evolution}
\end{figure*}

Let us construct a useful model for understanding the evolution of the system, under equations \eqref{eqn: evolution}.
We see from figure \ref{fig:example evolution} that $\mathbf{e}$ and $\mathbf{j}$ explore their available phase-space, with $\omega$ and $\Omega$ varying wildly from one encounter to the next. Indeed, if the angles $\omega$ and $\Omega$ change sufficiently fast, then these rapid (in comparison with the changes in $e$ and $i$) fluctuations may be approximated as random. This is of course only true on long time-scales, of the order $\tau_{\rm sec}$.
Equations~\eqref{eqn: evolution} then imply that the motion of $e$ and $i$ can be approximated as a partially-random walk, where, in each step, $\omega_n$ and $\Omega_n$ are essentially random, and the jumps are given by equations~\eqref{eqn: changes per encounter}. This walk is correlated in the sense that on short time-scales (the time it takes $\omega$ or $\Omega$ to scramble), $e$ and $i$ retain memory of their previous states. 

\subsection{External perturbations}
\begin{figure}
    \centering
    \includegraphics[width=0.45\textwidth]{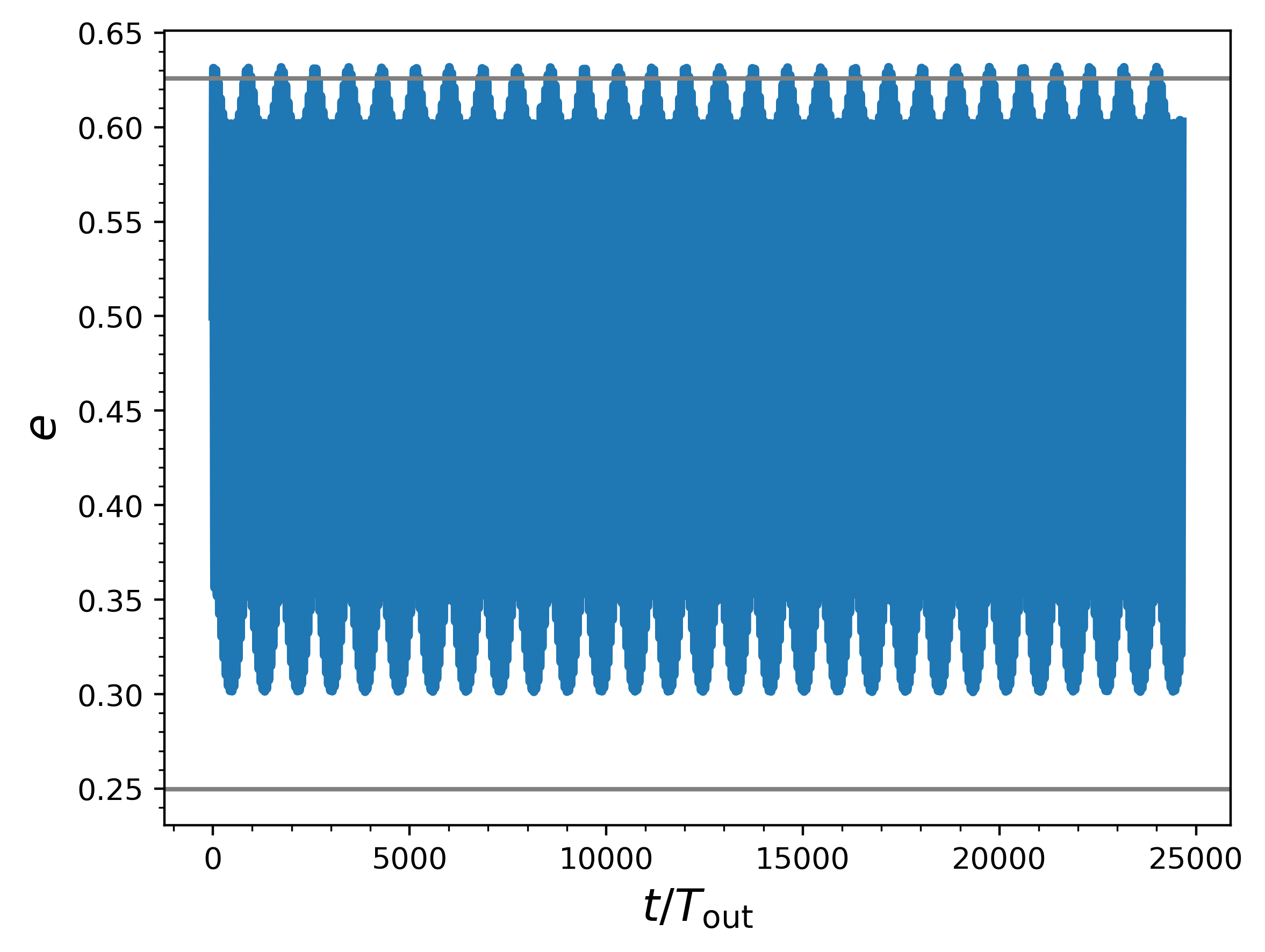}
    \caption{The eccentricity evolution for the example parameters: $m_1 = 1.9\, M_{\odot}$, $m_2 = 0.1\, M_{\odot}$, $m_3 = M_{\odot}$, $r_{\rm p} = 7a_{\rm in}$, $i_0 = 7\pi/9$, from {\tt rebound}.}
    \label{fig:e of t long term rebound}
\end{figure}

The true evolution of the triple does not display such an eccentricity drift. This is because the triple does not evolve with a random $\Omega$ and $\omega$ at every time-step. For example, at $\mathcal{O}(\eps)$, $j_z$ and the single-averaged Hamiltonian are conserved, meaning that the triple's evolution is integrable, leading to periodic motion. The {\tt rebound} simulation is also periodic, as can be seen in figure \ref{fig:e of t long term rebound}, which indeed does not display an eccentricity drift. We thus conclude that quasi-hierarchical triples of point particles, without general relativity, which are in isolation, exhibit periodic behaviour. 

However, it is interesting to ask if there are astrophysical systems where the triple's evolution does display a quasi-random eccentricity drift, e.g.~if there is an extra randomness on a time-scale $\tau_{\rm sec}$. This can occur, for example, in the context of relativistic phase-space diffusion \citep{Kunz2022,Hamilton2024b}, or if the outer orbit is itself subject to perturbations, which can change $i$ and $\Omega$, e.g.~from by-passing stars \citep{Hamersetal2018}, a fourth object \citep[e.g.,][figure 6]{Yangetal2025} or tri-axiality of the external potential \citep{Petrovich2017}, where it is known that the inner eccentricity can experience a drift to ever higher values and that eccentricities experience a diffusion. A related study by \cite{Winter_etal2024}, saw an enhancement of the probability of reaching very high eccentricities if both galactic tides and fly-bys perturb a binary, rather than just one of these processes.

There is a possibility that the outer angular momentum would shrink to such a value that $r_{\rm p, out}$ would cease to be larger than $\sim a_{\rm in}$, thereby violating the quasi-hierarchical assumption made in this paper. Indeed, if that occurs, the conclusions of this section would not apply. However, generically, the time-scale for the \emph{direction} of the outer orbit to diffuse is shorter than the time-scale for it to change its magnitude (e.g.~in the case of by-passing stars, leading to vector resonant relaxation being more rapid than its scalar counterpart; \citealt{KocsisTremaine2011,Hamersetal2018}); in that case, the outer angular momentum can change without violating this assumption.
For example, if $\Omega$ is subject to a random change in each outer orbit, then the map \eqref{eqn: evolution} will lead to a random walk of $\mathbf{e}$ and $i$. The bottom right panel of figure \ref{fig:example evolution} shows why that is: $\Omega$ essentially changes by an $\mathcal{O}(\pi)$ amount over one secular time, so over many secular times, its value (modulo $2\pi$) is essentially random; hence, if the true evolution randomises $\Omega$ over a secular time, the effect on the inner binary will be qualitatively the same as if $\Omega$ were randomised by the map (with its $\mathcal{O}(\eps^3)$ errors leading to non-periodic evolution). 

Let us now postulate that such a triple exists, where $\Omega$ experiences random perturbations on a $\tau_{\rm sec}$ time-scale. These can be more or less arbitrary, in-so-far as they remove any additional constraints that equations \eqref{eqn: evolution} do not satisfy, that the exact, isolated triple does. We therefore assume here, that if constraints on the angular-momentum direction of the outer orbit are removed---manifesting in, e.g., a random changes in $\Omega$, thought of as coupling the outer orbit to a bath with which it can exchange angular momentum---the triple evolves qualitatively as in equations \eqref{eqn: evolution}. Essentially, we take the outer orbit to be a different one each time-step, but with the same values of $r_{\rm p,out}$ and $i$. Thus, each time-step is essentially a new parabolic encounter. 

Let us focus on $e$ for such a hypothetical triple: its random walk has two boundaries $e_{\rm l} = 0$ and $e_{\rm u} = 1$ (this can be replaced by any value $e_{\rm u} = e_{\rm coll}$ above which a collision occurs, or the inner binary coalesces much more quickly than an outer orbital period). The first-order part of $\Delta e$ has zero average (over $\omega$ and $\Omega$), and its second moment is
\begin{equation}\label{eqn:rms delta e first order}
	\sqrt{\left\langle \Delta e^2 \right \rangle} = \frac{15\pi}{4\sqrt{2}}\eps e \sqrt{1-e^2}\sin^2 i + \mathcal{O}(\eps^2). 
\end{equation}
Thus, it appears that the walk would inevitably reach the boundary---either $e = e_{\rm u}$ (where it would terminate) or at $e = 0$ (where it stalls). 

There is another drift term, which can overcome the diffusion: consider equation~\eqref{eqn: eccentricity change second order}, when averaged over $\omega$ and $\Omega$, \emph{viz.}
\begin{equation}
	\begin{aligned}
		\left\langle \Delta e \right\rangle & = \frac{9\pi^2}{512}\eps^2 e \Big[4(81e^2-56)\cos 2i + (39e^2+36)\cos 4i \\ & 
		- 299e^2 + 124\Big].
	\end{aligned}
\end{equation}
At $i$ close to $0$ and $\pi$, the first-order diffusion \eqref{eqn:rms delta e first order} is negligible, and hence the eccentricity changes are dominated by this $\left\langle \Delta e \right\rangle$. For $i> \pi - \arctan\left(\frac{\sqrt{\sqrt{97}-1}}{2\sqrt{6}}\right)$, this drift is negative for all $e$, whence it seems that sufficiently retrograde orbits would consistently lose eccentricity under such an evolution.
We show in appendix \ref{appendix: random walk stuff} that this is indeed the case. The boundary between the region where $e$ can walk randomly to high eccentricities is given by requiring that $i_0$ be sufficiently close to $\pi/2$, for an orbital flip to be possible within one secular period, i.e.,
\begin{equation}
    \frac{1}{\eps} \Delta j_z \sim j_z\,,
\end{equation}
where $\Delta j_z$ is maximised over all values of $\omega$ and $\Omega$. 
If $i_0 -\pi/2 \ll 1$, this simplifies to 
\begin{equation}\label{eqn:i crit}
    i_0 \leq i_{\rm crit} \equiv \pi - \arccos \left(\min\set{1,\frac{300\pi\sqrt{1-e^2}\eps}{64(1-e^2)}}\right)\,.
\end{equation}
We verify this in appendix \ref{appendix: random walk stuff}.

Near the boundary $e=0$, we have $\Delta e \sim e$. This means that even if all changes $\Delta e_n$ are negative (an event with an exponentially small probability), $e$ cannot decrease faster than as a geometric sequence---it does not reach $e=0$ at a finite time. Hence, for $i_0 \leq i_{\rm crit}$ the walker, if starting at $e_0>0$, never reaches $e=0$ exactly. For $i_0 \geq i_{\rm crit}$, the negative drift eventually wins, causing $i$ to approach $\pi$ and $e$ to tend asymptotically to $0$. 

This behaviour is in contrast with the boundary $e_{\rm u}$, where equation \eqref{eqn: eccentricity change second order} yields $\Delta [1-e] \sim \sqrt{1-e}$, whence here, if steps do add up coherently, $e$ can reach $e=e_{\rm u}$ in a finite time.\footnote{This may be formalised using a \cite{Gronwall1919}-like estimate.} In conclusion, the boundary at $e=0$ is benign, and can never be reached at a finite time, while the boundary $e=e_{\rm u}$ can be. Therefore, the upper boundary, at $e = e_{\rm u}$, will inevitably be reached for $i_0 \leq i_{\rm crit}$, for every initial non-zero eccentricity. Observe that it is impossible for $\Delta e$ to be larger than $1-e$; this is both a property of equation \eqref{eqn: eccentricity change second order}, and an immediate consequence of orbit-averaging over the inner orbit. Such orbit-averaging was shown to be an excellent approximation as long as $r_{\rm p} \geq 3 a_{\rm in}$ \citep{HamersSamsing2019a,Samsingetal2019} for individual parabolic encounters. 

As we are agnostic here about the astrophysical environment that leads, in reality, to a randomisation of $\Omega$ or $\omega$, we leave a more detailed study of the long-term evolution under the map \eqref{eqn: evolution}, and its consequences, to appendix \ref{appendix: random walk stuff}. 

\section{Discussion and summary}
\label{sec:discussion}
In this paper, we analysed the dynamical evolution of quasi-hierarchical triples, where there is a clear hierarchy between the orbital times of the outer and the inner orbits, but the outer eccentricity is so high, that the outer orbit's time at pericentre is not as significantly larger than the inner orbital period. We modelled these systems using a map---from one outer pericentre to the next---where each step was approximated as a parabolic encounter between the inner binary and the tertiary companion. These steps were of course correlated, because it is the same tertiary that encounters the same binary. The orbital evolution displays secular oscillations of $\mathbf{e}$ and $\mathbf{j}$, qualitatively similar to ZLK oscillations. However, the orbit is sometimes capable of reaching an even larger maximum eccentricity, or a lower one, than what is predicted by pure ZLK evolution. These deviations are present at second-order quadrupole order, occurring on a quadrupole time-scale, even for equal masses, where the octupole vanishes. We showed that such a modelling agrees with direct three-body integration over secular times.

Over many secular times, the map \eqref{eqn: evolution} accrues $\mathcal{O}(\eps^3)$ errors, which cause it to deviate from numerical integration. Instead, that map causes the angles $\omega$ and $\Omega$ to fluctuate significantly during these kicks, so the evolution of $e$ and $i$ can be approximated as a random walk. This behaviour is not present in isolated quasi-hierarchical triples, but we hypothesise that it can arise if the outer orbit is not forced to conserve its angular momentum---if the outer orbit is weakly coupled to a system with which is can exchange angular momentum. This implies that, e.g., $\Omega$ is randomised between kicks, so the inner orbit's evolution is that same as what would happen under consecutive interactions with \emph{different} outer, parabolic orbits. 

\emph{If} indeed the map \eqref{eqn: evolution} is a reasonable approximation in that case, we showed that this walk inevitably leads to $e$ reaching arbitrarily large values, and that it necessarily eventually reaches the boundary $e=1$. The time to do so was found in appendix \ref{appendix: random walk stuff} to be the diffusion time-scale $t_{\max} = \tau f\left(e_0,i_0|e_{\rm u}\right)$, which we showed scaled like $\eps^{-2}$. We leave a quantitative study of the applicability of the random-walk model to non-isolated triples for future work. 

Applications for this theory include---but are not limited to---an enhancement of the triple channel from gravitational-wave source formation. For isolated triples, the time it takes the inner binary to coalesce is modified by an additive factor $\delta_{\max}$, plotted in figure \ref{fig:delta gw}. For triples where the map \eqref{eqn: evolution} is a faithful qualitative description of their long-term evolution, the diffusion process speeds up the time to reach high eccentricities significantly, driving the inner binary all the way to small pericentres, where relativistic precession and/or emission become stronger than quasi-hierarchical evolution. In systems which obey the hierarchy \eqref{eqn:fundamental assumption} but not the stronger condition \eqref{eqn:outer pericentre upper bound strong quasi-hierarchy}, one must account both for the quasi-hierarchical `jumps' due to outer-pericentre passages, and for the deterministic secular contribution of the rest of the outer orbit, on $\mathcal{O}(T_{\rm out}/\eps^2)$ times. That is, the map \eqref{eqn: evolution} must be modified to include, e.g.~the changes in $\mathbf{e}$ and $\mathbf{j}$ induced by the ZLK effect, in addition to those in \eqref{eqn: changes per encounter}. This combined treatment is left for future work.

\section*{Acknowledgements}
We are grateful to Francesco Mori for helpful discussions on random walks, to Evgeni Grishin and Mor Rozner for very helpful comments on the manuscript, and to Samuel Macguire and Evgeni Grishin for sharing their implementation of the \cite{HamersSamsing2019a} equations in \cite{Macguireetal2026}. We thank Mor Rozner for her encouragement to finish the paper. We thank the anonymous referee for their helpful comments, which improved the manuscript.
This work was supported by a Leverhulme Trust International Professorship Grant (No.~LIP-2020-014). The work of Y.B.G.~was partly supported by a Simons Investigator Award to A.A.~Schekochihin (No.~930121). J.~Samsing's work was partly supported by the ERC Starting Grant No.~121817-{BlackHoleMergs} (PI: Samsing) and by the Villum Foundation (Villum Fonden; grant No.~29466). The Centre of Gravity at the Niels Bohr Institute is a Centre of Excellence funded by the Danish National Research Foundation under grant No.~184.

\section*{Data Availability}
Simulations conducted for this article utilised the publicly available code \verb"rebound" \citep{ReinLiu2012,ReinTamayo2015}.\footnote{\url{https://rebound.readthedocs.io/}} Scripts and input files used to make the figures in this article will be shared upon reasonable request to the corresponding author.




\bibliographystyle{mnras}
\bibliography{triples}


\appendix

\section{Some details on quasi-hierarchical scattering}
\label{appendix: details quasi hierarchical}

In this appendix we discuss some features of the quasi-hierarchical interaction between the inner binary and the third star. We also test the approximation by comparing it with the results of a simple three-body simulation. 

\subsection{Magnitude of eccentricity change and angular-momentum flipping}
\begin{figure*}
	\centering
	\includegraphics[width=0.32\textwidth]{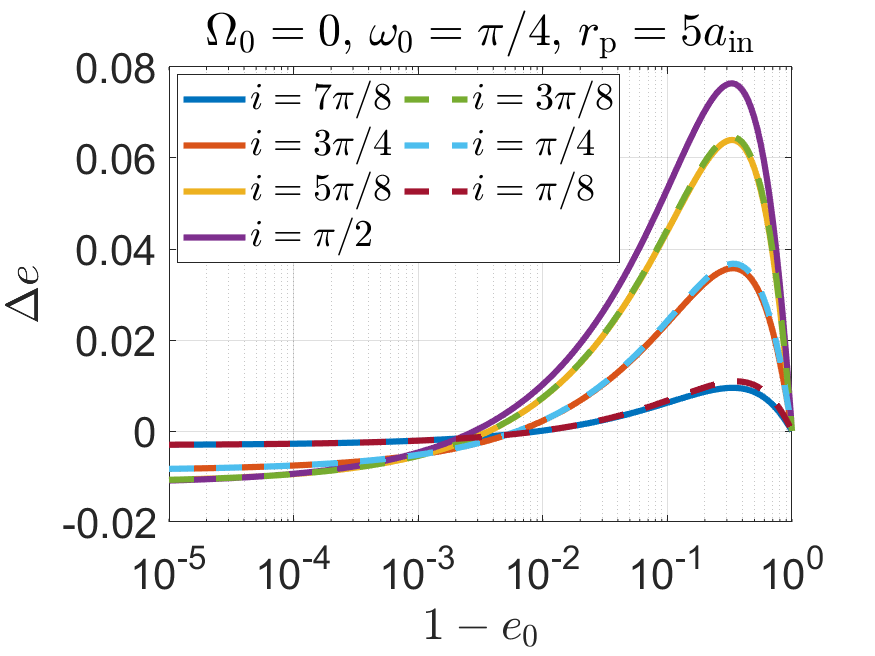}
	\includegraphics[width=0.32\textwidth]{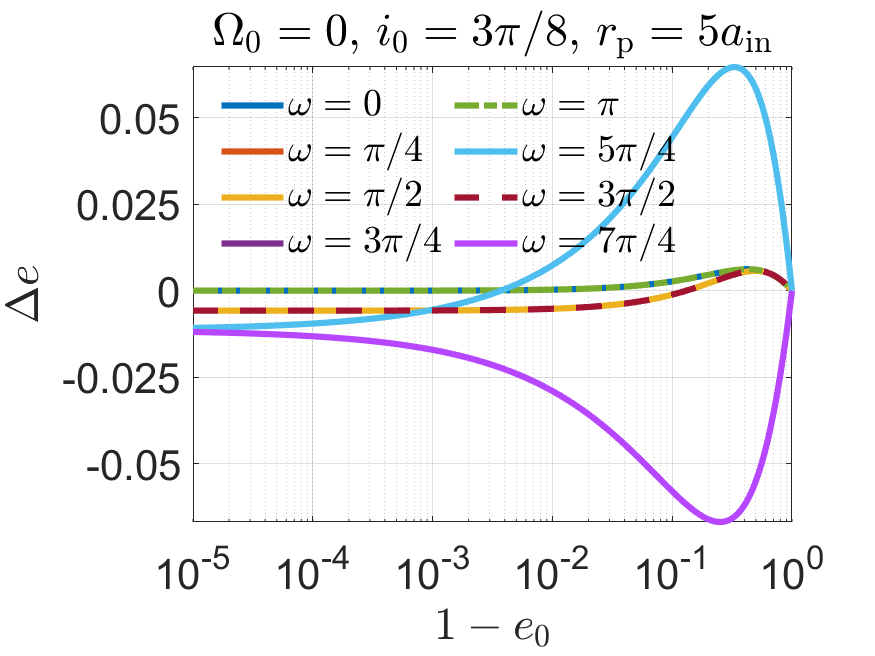}
	\includegraphics[width=0.32\textwidth]{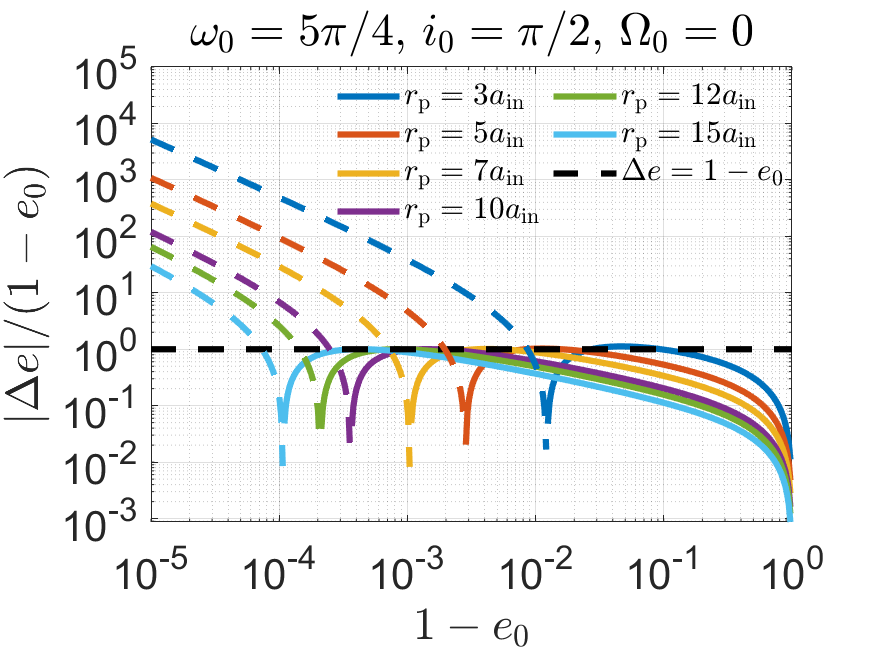}
	\caption{The change $\Delta e$ in eccentricity, given by equation~\eqref{eqn: eccentricity change second order}, as a function of the initial eccentricity $e_0$, for various values of some of the parameters of the problem. \emph{Left}: varying the inclination between the inner and outer orbits; \emph{middle}: varying the inner orbit's argument of pericentre; \emph{right}: varying $\eps$. The dashed lines on the \emph{right} panel correspond to $\Delta e < 0$, while the full lines are $\Delta e\geq 0$.}
	\label{fig:delta e of e}
\end{figure*}

The map \eqref{eqn: changes per encounter}, applied once, yields a positive or negative eccentricity change, depending on the angles $i,\omega, \Omega$ and on $r_{\rm p}/a_{\rm in}$. In figure \ref{fig:delta e of e} we plot $\Delta e$ as a function of $1-e_0$, after one outer pericentre, for various values of these parameters. Observe, that $\Delta e \leq 1- e_0$.

\subsection{Orbit flipping}
A single encounter with the tertiary can also result in $i$ crossing $\pi/2$. However, this occurs when $\Delta \mathbf{j}$ is as large as $\mathbf{j}$, and therefore is not immediately captured by equations \eqref{eqn: changes per encounter}. To account for it, we first check whether 
\begin{equation}\label{eqn:flipping criterion}
    \sqrt{1-e^2} < \frac{15\eps\pi e^2}{4}\,;
\end{equation}
if so, then we evolve $e$ via equation \eqref{eqn: eccentricity change second order} as usual; but for the angles, we first evolve $i$, $\Omega$ and $\omega$ according to the first-order part of equation \eqref{eqn: changes per encounter} (truncated at first order), and then insert those new values into the second-order expressions \eqref{eqn: changes per encounter}, to obtain $i$, $\Omega$ and $\omega$ for the next round. 
If inequality \eqref{eqn:flipping criterion} is not satisfied, we evolve the orbital parameters as stated in the main text.

\begin{figure}
	\centering
	\includegraphics[width=0.48\textwidth]{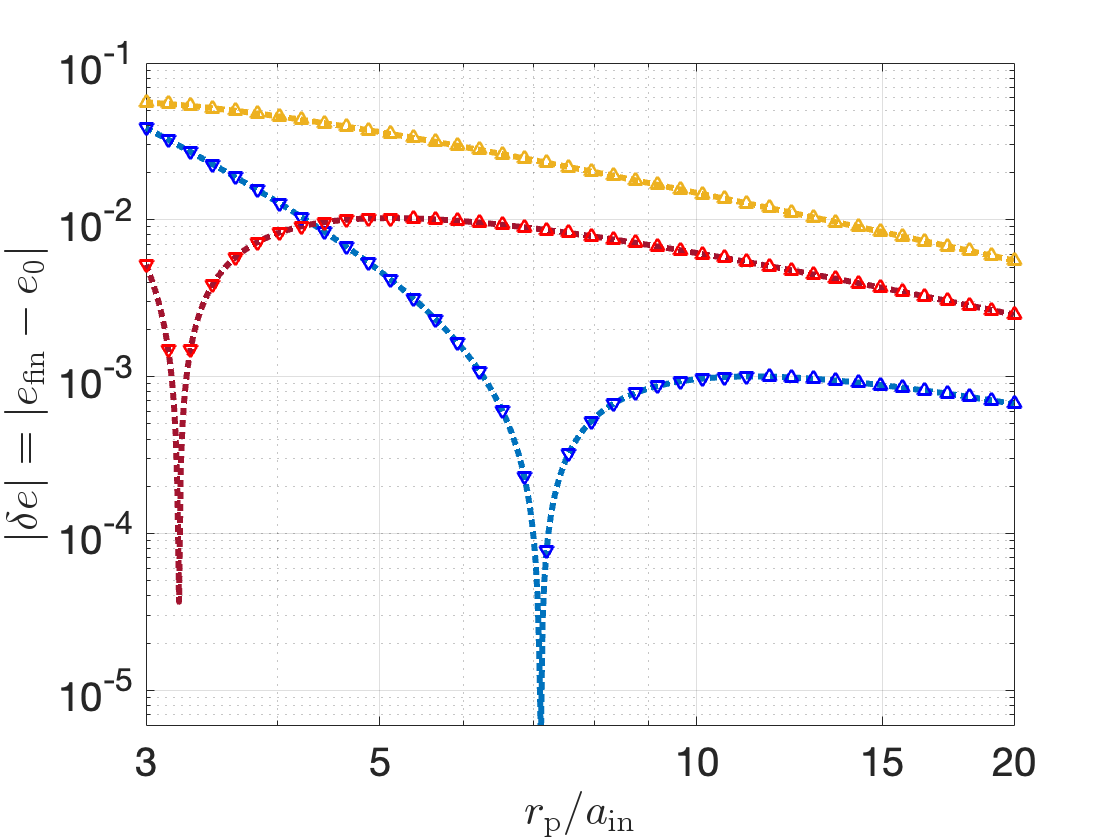}
	\caption{The eccentricity change given by equation \eqref{eqn: eccentricity change second order}, for $m_1 = m_2 = m_3 = 20\,M_\odot$, $i_0 = \pi/2$, $\Omega = 0$, $\omega = \pi/4$, and various eccentricities: $e_0 = 0.95$ (yellow), $0.99$ (red) and $0.999$ (blue). The colours and initial conditions are chosen for comparison with \citet[][figure 4, top panel]{Samsingetal2019}. Upward-pointing triangles denote $i_{\rm final} \leq \pi/2$, while upside-down triangles denote flipped systems, with $i_{\rm final} > \pi/2$.}
	\label{fig:Samsing2019}
\end{figure}

This procedure ensures that orbital flipping is correctly accounted for; that this is the appropriate is evident from figure \ref{fig:Samsing2019}, which reproduces the correct flipping results from the three-body simulations of \cite{Samsingetal2019}: this figure should be compared with the top panel of figure 4 of that work, where the orbit flipping was marked based on three-body simulations; the prescription described above reproduces that. 

\subsection{Phase-space constraints}
\label{appendix:e and j constraints}
The vectors $\mathbf{e}$ and $\mathbf{j}$ only parameterise a $4$D phase-space, spanned by $e$ and the three angles $i,\omega$ and $\Omega$, so they satisfy two scalar constraints:
\begin{align}
	& e^2 + j^2 = 1\,, \\ &
	\mathbf{e}\cdot\mathbf{j} = 0\,.
\end{align}
The implementation \eqref{eqn: changes per encounter} does not necessarily preserve them, exactly. We follow the method proposed by \cite{Macguireetal2026} in ensuring that they are preserved exactly. In each step, we used equations \eqref{eqn: changes per encounter} to advance $\mathbf{e}$ and $\mathbf{j}$. If the new $e^2 + j^2 \neq 1$ or $\mathbf{e}\cdot \mathbf{j} \neq 0$, we keep the new value of $\mathbf{e}$ and one component of $\mathbf{j}$, but we solve the two constraint equations for the two other components of $\mathbf{j}$. This introduces $\mathcal{O}(\eps^3)$ changes in just the right way to preserve the constraints. In all cases but $1-e \sim \mathcal{O}(\eps^3)$, there exists a component $j_i\in\set{j_x,j_y,j_z}$ such that the two constraint equations have a real solution for the two other components. We remark that while the improvement of \cite{Macguireetal2026} is equivalent to the \cite{HamersSamsing2019a} map over a single iteration, it yields a dramatic improvement of the agreement with \texttt{rebound} over many time-steps.

\subsection{Angular-momentum conservation}
\label{appendix:angular momentum conservation}

The random walk described in this paper implies that the map \eqref{eqn: evolution} tends to bring the inner eccentricity $e$ to higher values, so that the magnitude of the inner orbit's angular momentum decreases. This angular momentum is transferred to the outer orbit, and the na\"{i}ve map \eqref{eqn: evolution} does not readily account for that. To conserve the total angular momentum, we update $r_{\rm p}$ in each iteration, so that the total angular momentum is conserved. Having modified $\mathbf{j}$ (and thus $\mathbf{J}_{\rm out}$, the outer angular momentum), we rotate the frame of reference so that in the next iteration of equation \eqref{eqn: evolution}, $\mathbf{J}_{\rm out} \parallel \zhat$; we update $i$ and $\Omega$ accordingly.

\section{Long-term evolution under second-order map}
\label{appendix: random walk stuff}

In this appendix we discuss several properties of the long-term evolution under the map described in \S \ref{sec:quasi-hierarchical triples}. We stress that, as remarked in \S \ref{sec:external}, it is a poor description of isolated quasi-hierarchical triples over time-scales much longer than {(a few)$\times \tau_{\rm sec}$}. However, it may apply qualitatively for triples whose outer orbit can exchange angular momentum with a reservoir. The map itself displays interesting properties, so we keep this discussion here. 

\begin{figure*}
	\centering
	\includegraphics[width=0.49\linewidth]{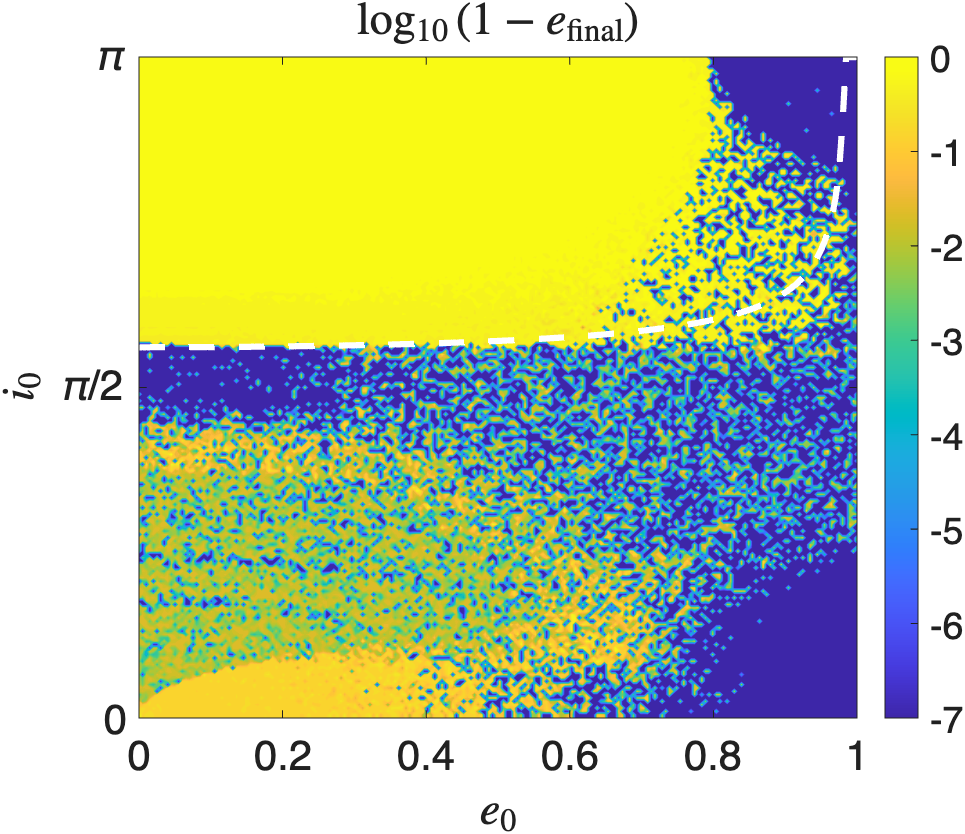}
	\includegraphics[width=0.49\linewidth]{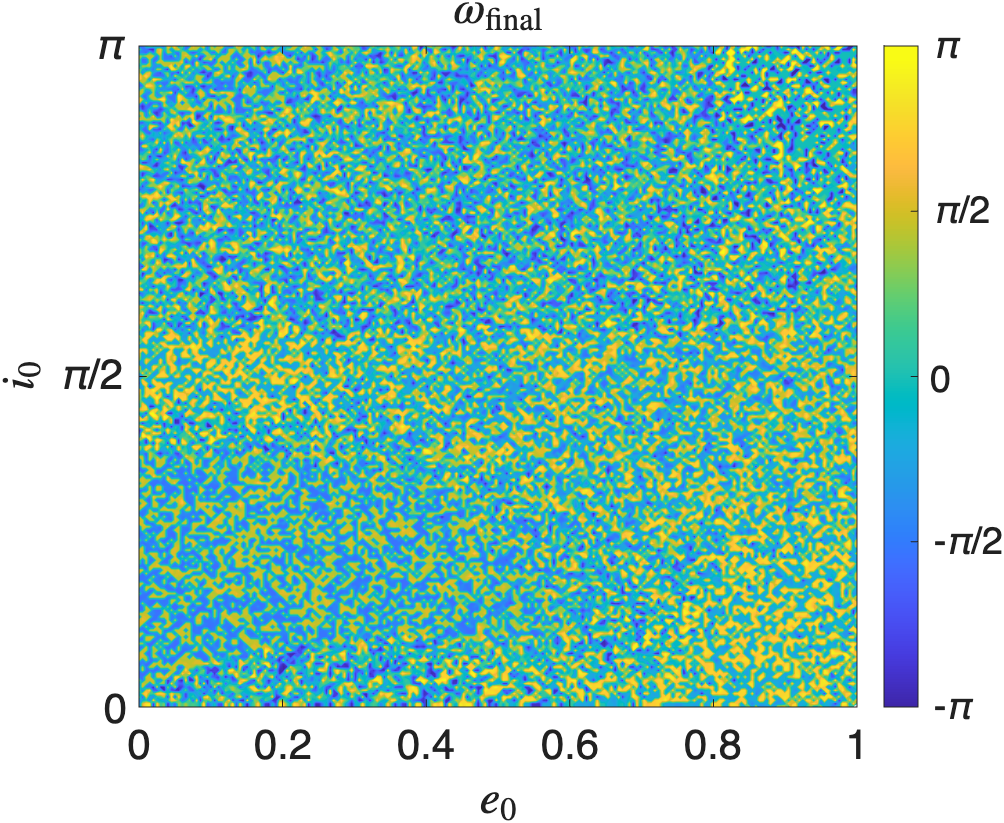}
	\includegraphics[width=0.49\linewidth]{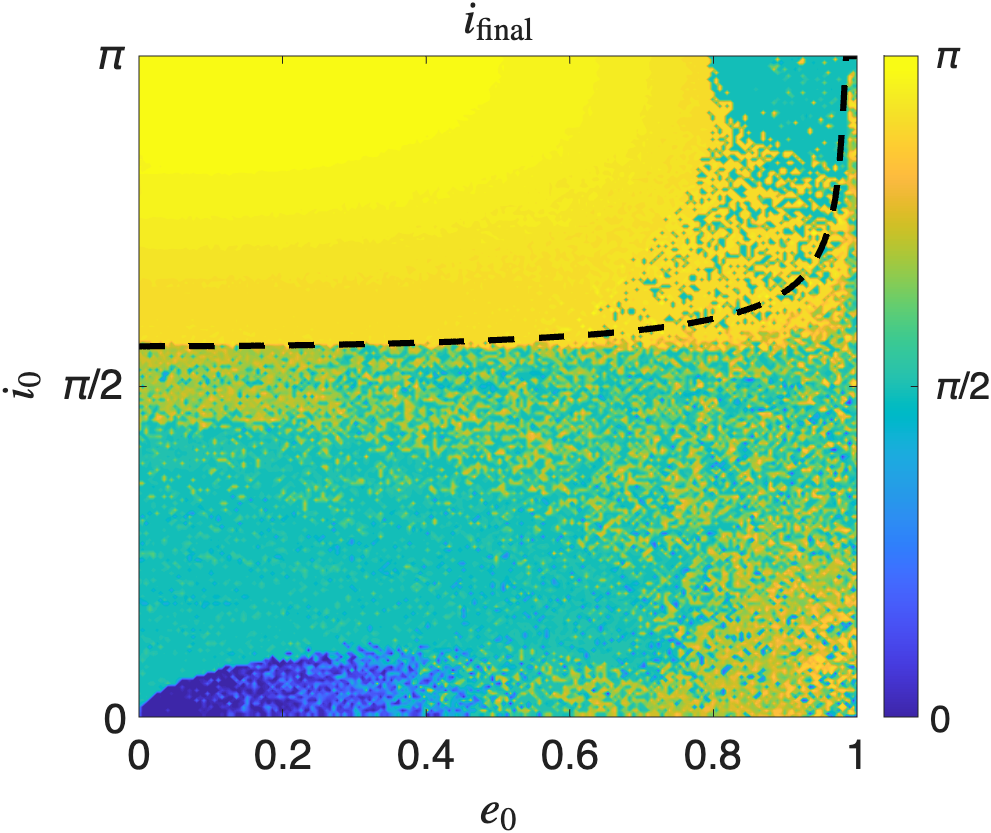}
	\includegraphics[width=0.49\linewidth]{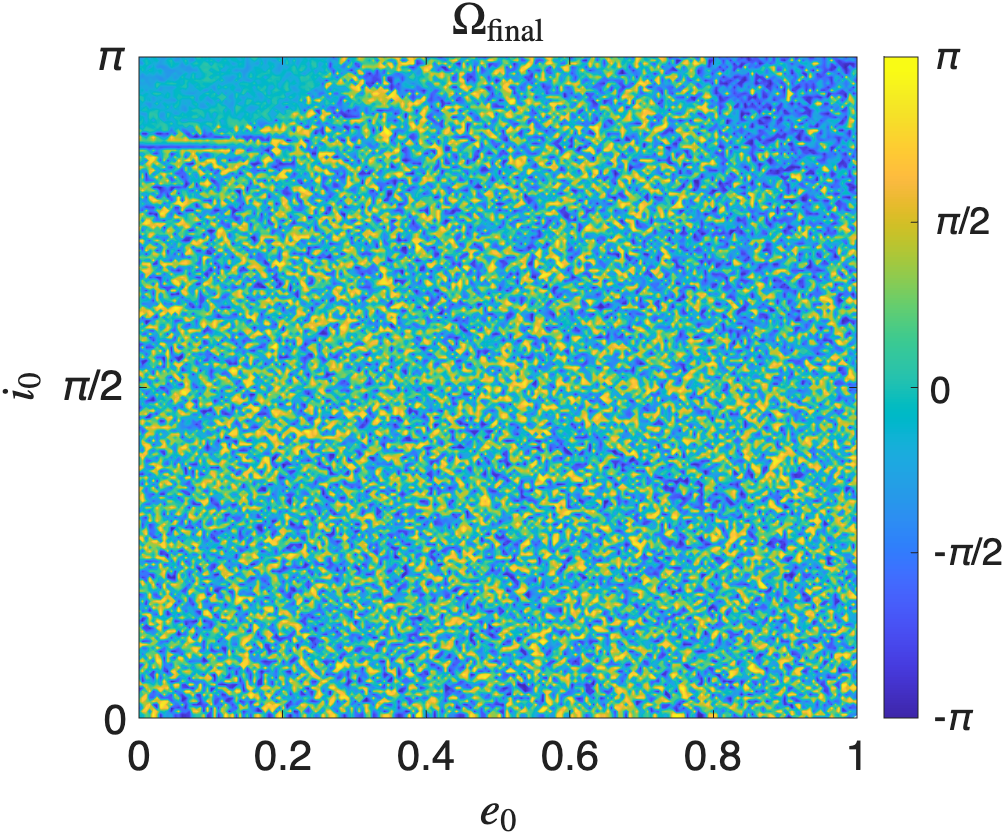}    
	\caption{The values of orbital parameters after $10000$ outer orbits, for $r_{\rm p} = 5a_{\rm in}$, of randomly chosen initial $\omega$ and $\Omega$, as functions of the initial eccentricity $e_0$ and inclination $i_0$, for an equal-mass triple. The system is frozen when $e_n$ reaches $1-10^{-7}$. One can see some spin flips in the \emph{bottom left} panel. $\mathbf{e}$ and $\mathbf{j}$ are evolved using equations~\eqref{eqn: evolution}, accounting explicitly for angular-momentum conservation. The dashed curve is $i_{\rm crit}$ from equation \eqref{eqn:i crit}.}
	\label{fig:phase diagram}
\end{figure*}

Figure \ref{fig:phase diagram} shows the value of the orbital parameters after $10000$ orbits of the outer binary, as functions of the initial condition $e=e_0$, $i=i_0$ (the angles were randomly selected for each initial condition), for $r_{\rm p} = 5a_{\rm in}$.  
We see that given time, more and more initial conditions reach $e_{\rm u}$ if the initial condition is below $i_{\rm crit}$, given by equation \eqref{eqn:i crit}, and they tend to $e = 0$ and $i=\pi$ if it is above, except for very high initial eccentricities, where the diffusion can nonetheless overcome the negative drift. 

\subsection{First-passage time}

\begin{figure*}
	\centering
	\includegraphics[width=0.314\textwidth]{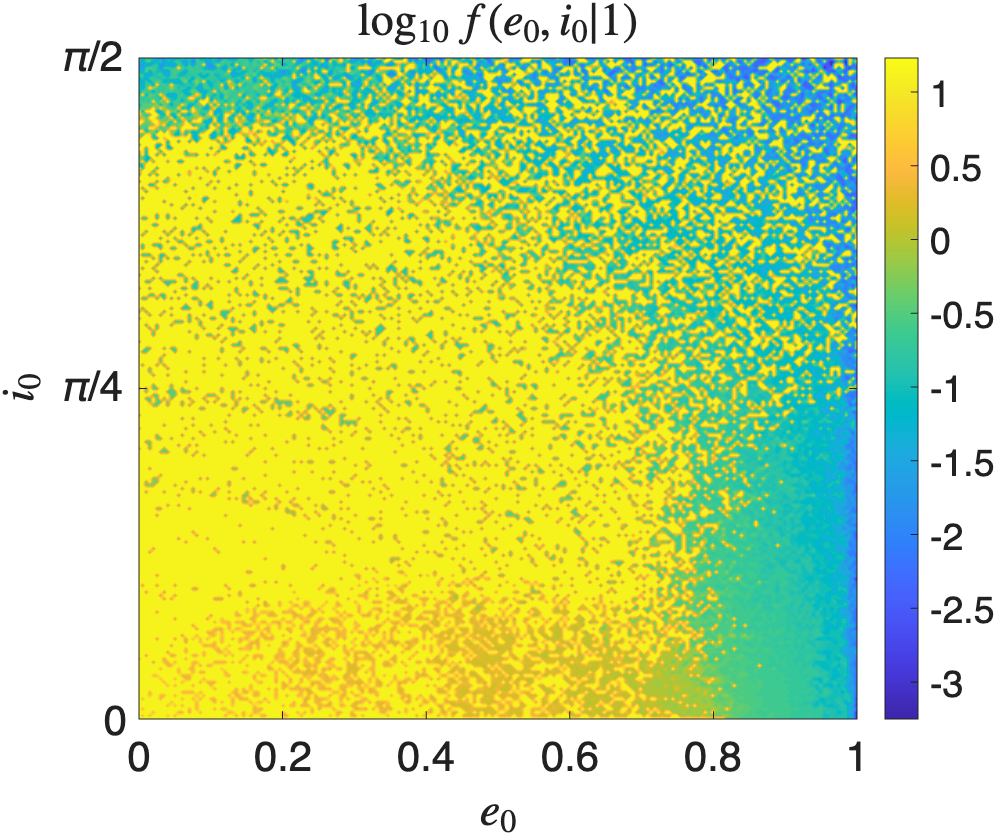}
	\includegraphics[width=0.339\textwidth]{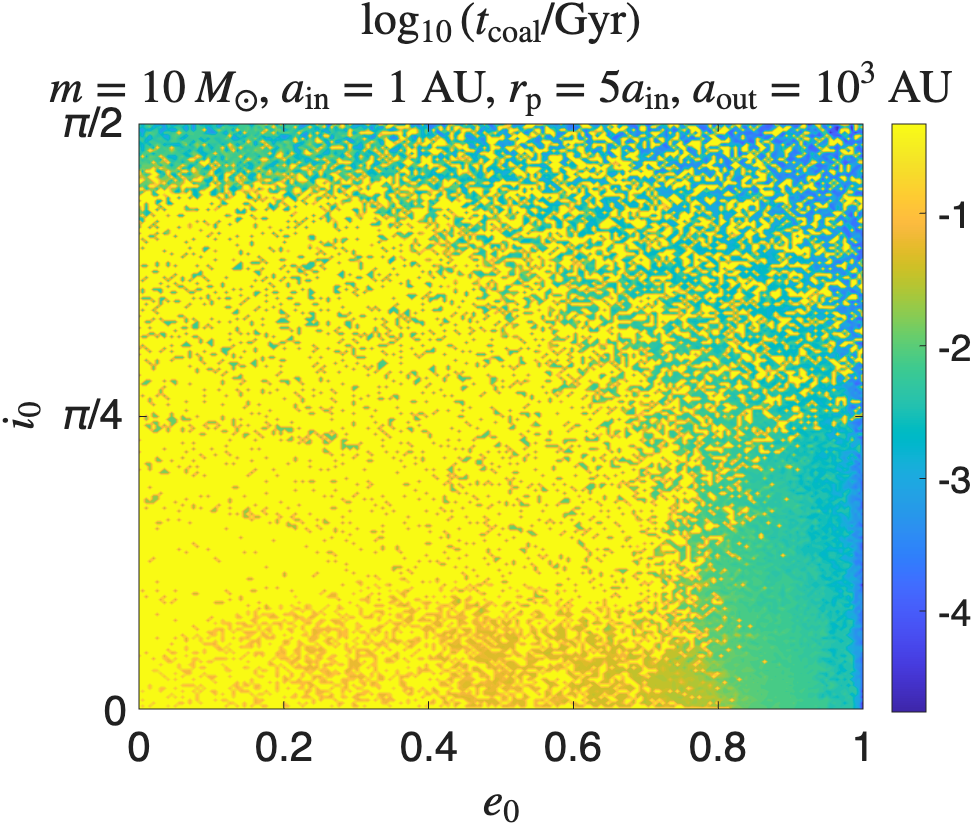}
    \includegraphics[width=0.302\textwidth]{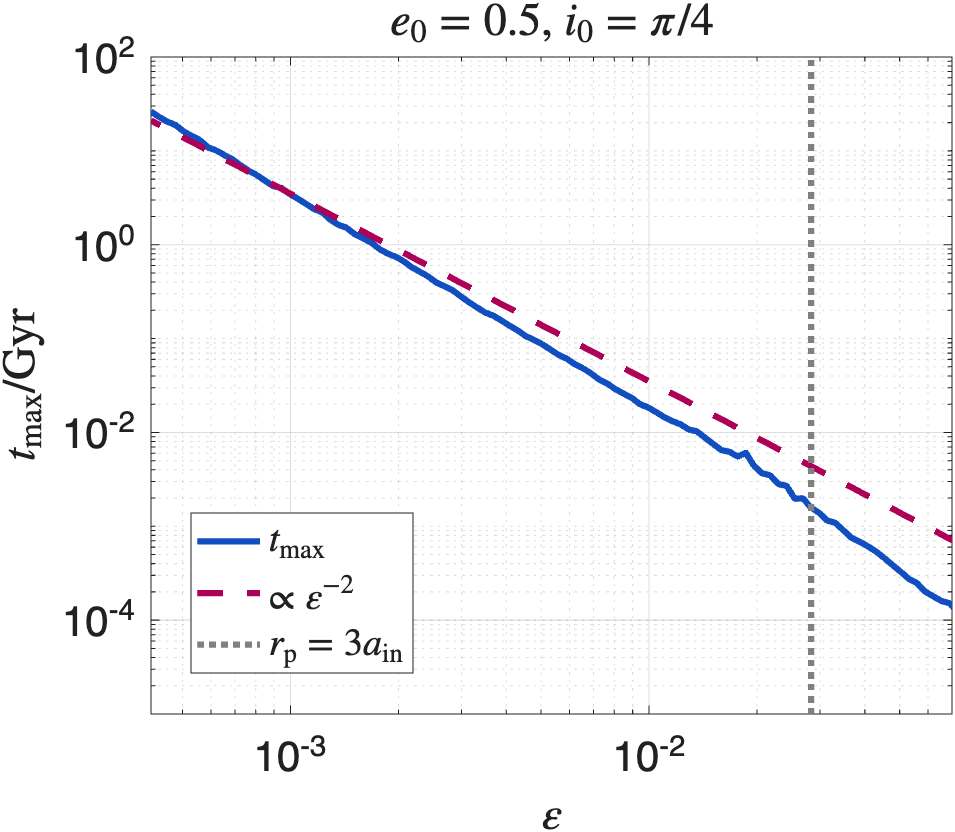}
	\caption{\emph{Left} panel: the function $f(e_0,i_0|1)$, defined in equation~\eqref{eqn:diffusion time-scale}. We only show $i_0 \leq \pi/2$ to so that the $i_0 < i_{\rm crit}$ (equation \eqref{eqn:i crit}) and a diffusion to $e_{\rm u}$ is ensured. \emph{Middle}: $t_{\max}$, measured directly from evolving the triple (as in figure \ref{fig:phase diagram}), for equal-mass triples with $m=10\,M_\odot$, $a_{\rm in} = \textrm{AU}$, $a_{\rm out} = 1000\; \textrm{AU}$, $r_{\rm p} = 5a_{\rm in}$. \emph{Right}: $t_{\max}$ for an example value of $(e_0,i_0)$ and random $\omega_0$ and $\Omega_0$ (uniform in $[0,2\pi]$, averaged over multiple draws), for the parameters of the \emph{middle} panel, except $r_{\rm p}$, which is varied.}
	\label{fig:t coal}
\end{figure*}

For $i_0 \leq i_{\rm crit}$, the time to reach $e_{\rm u}$ (see \S \ref{sec:external} above) may be obtained from standard first-passage-time calculations, which we describe in appendix \ref{appendix: first passage}. If the initial values of $\omega$ and $\Omega$ are selected at random, then the mean time to reach $e=e_{\rm max}$ is given by
\begin{equation}\label{eqn:diffusion time-scale}
	t_{\max} = T_{\rm out} \frac{f(e_0,i_0|e_{\rm u})}{\eps^2} \equiv \tau f(e_0,i_0|e_{\rm u}) \; ,
\end{equation}
where the diffusion time-scale is defined as $\tau \equiv T_{\rm out}/\eps^{2}$, and $f$ is independent of the semi-major axes and $\eps$, and only depends on the initial condition $(e_0,i_0)$ and on $e_{\rm u}$, and satisfies the boundary conditions $f(e_{\rm u},i_0|e_{\rm u}) = 0$, $f(0,i_0|e_{\rm u}) = \infty$. We plot $f$ and $t_{\max}$ in figure \ref{fig:t coal} for $i_0 \leq \pi/2 < i_{\rm crit}$; $f$ is measured for each pair $(e_0,i_0)$ as the time to reach $e_{\rm u}$, where $\omega_0$ and $\Omega_0$ are chosen randomly, evolved using equations~\eqref{eqn: evolution}. $f$ is noisy in the left panel of figure \ref{fig:t coal}, because it measured from one realisation of $(\omega_0,\Omega_0)$; however, the fact that it is of order unity throughout the parameter space lends credence to the random-walk model. It is evident from the right panel of figure \ref{fig:t coal}, that $t_{\rm max}$ follows the $\eps^{-2}$ scaling predicted by this model to a good accuracy (except possibly at $r_{\rm p}$ near the limit $\sim 3a_{\rm in}$ where the theory breaks down).

Using equations~\eqref{eqn: hierarchy parameter definition} and \eqref{eqn:diffusion time-scale}, the maximum value of $r_{\rm p}$ for $t_{\max} \leq t_0$ for some $t_0$ is therefore 
\begin{equation}\label{eqn: maximum rp}
	r_{\rm p,max}^2 (t_0)\equiv \min\set{\left[\frac{t_0}{T_{\rm out}}\right]^{\frac{2}{3}}   \frac{m_3^{4/3}\left(m_{\rm b}M\right)^{-2/3}a_{\rm in}^2}{\left[ f(e_0,i_0|e_{\rm u})\right]^{2/3}},a_{\rm in}a_{\rm out}},
\end{equation}
where the requirement $r_{\rm p,max} \leq \sqrt{a_{\rm in}a_{\rm out}}$ stems from imposing a \emph{strong} quasi-hierarchical approximation, that is, that the change in $\mathbf{j}$ and $\mathbf{e}$ really be dominated by the pericentre. This is satisfied when $\eps(a_{\rm in},r_{\rm p}) \geq \eps(r_{\rm p},a_{\rm out})$, i.e.~$r_{\rm p}^2 \leq a_{\rm in}a_{\rm out}$, or, equivalently, $1-e_{\rm out} \leq a_{\rm in}/r_{\rm p}$.
Ideally, one would like the inequality to be a strong one, and this is the condition we require here:
\begin{equation}\label{eqn:outer pericentre upper bound strong quasi-hierarchy}
1-e_{\rm out} \ll \frac{a_{\rm in}}{r_{\rm p}}\, .
\end{equation}

\subsection{Comparison with secular effects}
\label{subsec:secular}
In contrast to secular effects, where the change in orbital parameters is minute over one outer orbit, here $\Delta e$ and $\Delta \mathbf{j}$ are non-negligible on the time-scale of one outer orbit; so the mechanism described in this paper dominates these time-scales. On longer time-scales, one potentially needs to consider the contributions of the entirety of the outer orbit---not just its pericentre. These are expected to matter over secular time-scales; for a hierarchical triple, the relevant one is the ZLK time-scale, for a corresponding circular outer orbit, given by \citep{Naoz2016}\footnote{If we had included the factor of $\left(1-e_{\rm out}^2\right)^{3/2}$ in $\tau_{\rm ZLK}$ in \citet[][equation~27]{Naoz2016}, this would have given $\tau_{\rm ZLK} \sim \tau_{\rm sec} \sim \tau$ by definition. The shortening of the secular time-scale in that case is artificial, because here we wish to compare the effect of the pericentre (encapsulated by $\tau$) to the secular effect of the rest of the outer orbit, \emph{but not including its pericentre} (encapsulated by $\tau_{\rm ZLK,circ}$).}
\begin{equation}
	\tau_{\rm ZLK,circ} = T_{\rm out} \sqrt{\frac{m_{\rm b}M}{m_3^2}\frac{a_{\rm out}^3}{a_{\rm in}^3}}\;.
\end{equation}
In contrast, the diffusion time-scale \eqref{eqn:diffusion time-scale} (which measures the effects of the quasi-hierarchical pericentre `kicks') is $\tau = T_{\rm out}/\eps^2$.
Comparing the two we find
\begin{equation}
	\frac{\tau_{\rm ZLK,circ}}{\tau} = \frac{m_3^2}{m_b^{1/2}M^{3/2}}\left(\frac{\sqrt{a_{\rm in}a_{\rm out}}}{r_{\rm p}}\right)^3 = \sqrt{\frac{m_{\rm b}}{M}} \frac{\eps^2}{\alpha^{3/2}} \; ,
\end{equation}
where $\alpha \equiv a_{\rm in}/a_{\rm out}$;
thus, if the quasi-hierarchy is strong enough, $\tau$ can be shorter than $\tau_{\rm ZLK,circ}$, and secular effects would be suppressed over the entire diffusion process. The quasi-hierarchical assumption \eqref{eqn:fundamental assumption} does allow for regimes where this is not necessarily the case, but it is guaranteed by the restriction \eqref{eqn:outer pericentre upper bound strong quasi-hierarchy}. 
To summarise, we have 
\begin{equation}
    \alpha^{3/4}\left(\frac{m_{\rm b}}{M}\right)^{1/4} \ll \eps \ll 1\,,
\end{equation}
to be in the strong quasi-hierarchical regime; and additionally, to reach $e_{\rm u}$ within a time $t_0$, from equation~\eqref{eqn:diffusion time-scale} 
\begin{equation}
    \eps \geq \eps_{\max}(t_0) \equiv  \sqrt{\frac{T_{\rm out}f(e_0,i_0|e_{\rm u})}{t_0}}\,.
\end{equation}

\subsection{Collision Probability}
\label{sec:probability}

No triple is an island, and the outer tertiary should experience perturbations from its environment, at the same rate that a binary with semi-major axis $a_{\rm out}$ would \citep{MichaelyPerets2019,Samsingetal2019,MichaelyPerets2020,GrishinPerets2022,Stegmann2024}. While \cite{MichaelyPerets2020}, for example, consider $e_{\rm out}$ growing to such a high value that $r_{\rm p} \sim a_{\rm in}$ (whereupon a non-hierarchical binary-single encounter ensues), the triple would become quasi-hierarchical, well before reaching that value. At these high eccentricities, we would expect $r_{\rm p}^2$ to be uniformly distributed after the tertiary is perturbed; this implies that it is much more likely that a quasi-hierarchical triple would form, than a fully non-hierarchical one (cf.~table \ref{tab:regimes}). If the triple does enter a democratic state, this would eventually result in the ejection or collision of one of the three stars \citep[e.g.,][]{Saslawetal1974,Hills1980,Arnoldetal2006,Manwadkaretal2020,GinatPerets2021a}, and 
the eccentricity distribution of the remnant binary after the encounter would be slightly super-thermal \citep{StoneLeigh2019,Samsingetal2022,GinatPeretes2023,Randoetal2025}; in the more probable quasi-hierarchical case, the inner orbit can even reach higher eccentricities.

Thus, hierarchical triples are expected to become quasi-hierarchical on the time-scale on which external perturbations would induce an order-unity change in $e_{\rm out}$. The outer orbit may be modified either by encounters from external perturbers, or by the Galactic tide. Fortuitously, the time-scales are quite similar \citep[e.g.,][]{HeislerTremaine1986,Binney,GrishinPerets2022,Hamilton2022,HamiltonModak2024}, and are collectively given by \citep{Samsingetal2019} 
\begin{equation}\label{eqn:external perturbation time-scale}
	\tau_{\rm ext} = \frac{1}{2 \pi G} \frac{\sigma}{\rho a_{\rm out}}\frac{m_*}{\left(M + m_*\right)},
\end{equation}
where $m_*$ is the average stellar mass of the triple's environment. We remark that $\tau_{\rm ext}$ is typically significantly longer than $\tau$; this can be seen in figure \ref{fig:t coal}, which shows that $\tau \ll \textrm{Gyr}$, while $\tau_{\rm ext} \sim \textrm{Gyr}$.

Let us, therefore, assume that $e_{\rm out}$ is thermally distributed. Then, the probability for the inner binary reaching $e_{\rm u}$ within a time $t_0$ is equal to the probability that $r_{\rm p} \leq r_{\rm p,max}(t_0)$, i.e.
\begin{equation}\label{eqn:probability}
	\begin{aligned}
        P(t_0|a_{\rm in},a_{\rm out},e_0,i_0) = 1-\left[1 - \frac{r_{\rm p,max}(t_0)}{a_{\rm out}}\right]^2 \approx 2\frac{r_{\rm p,max}(t_0)}{a_{\rm out}} \,.
	\end{aligned}
\end{equation}
This probability corresponds to stating that the outer eccentricity $e_{\rm out}$ is constantly excited to values which are thermally distributed by the environment, on a time-scale $\tau_{\rm ext}$. 

Given a model for hierarchical triples---for $e_0$,$i_0$, $a_{\rm out}$---one can use equation~\eqref{eqn:probability} to derive a probability $P(t_0|a_{\rm in})$ for reaching $e_{\rm u}$ within a time $t_0$. Furthermore, if $e_{\rm u}$ and $t_0$ are chosen such that the diffusion time plus the time-to-coalescence is less than the typical time between encounters with the outer body, one can finally find a probability of merging as a function of $a_{\rm in}$ and the masses, only. 

\subsection{Eccentricity boundary}
We determine $e_{\rm u}$ by requiring two separate criteria, which we described in the following two sub-sections.

\subsubsection{Maximum eccentricity reachable by random walk}
The random walk may only drive $e$ efficiently until a certain value of $e$, above which most values of $\omega$ (and the other angles) would yield $\Delta e < 0$ (cf.~figure \ref{fig:delta e of e}). Again, assuming $i\approx \pi/2$, we find that if $1-e\ll 1$, 
\begin{equation}
	\Delta e \simeq \frac{15 \pi }{4} \eps \sqrt{1-e^2}   \sin 2 \omega +\frac{225}{64} \left(\pi ^2 \eps^2 \cos 4 \omega -\pi ^2 \eps ^2\right)\; ;
\end{equation}
for $\Delta e \geq 0$, one must have 
\begin{equation}
	\sqrt{1-e_{\rm u}^2} \geq \frac{15\pi}{8}\eps\sin 2\omega .
\end{equation}
Squaring and averaging this over $\omega$ one finds 
\begin{equation}\label{eqn:emax from random walk}
	1-e_{\rm u}^2 \geq \frac{225\pi^2}{128}\eps.
\end{equation}

\subsubsection{Beyond-Newtonian point-particle effects}

A cut-off for $e_{\rm u}$ can also be derived by requiring that it not be sufficiently high, that the orbital parameters of the inner binary would evolve significantly over one outer orbit, e.g. due to gravitational-wave emission, tidal effects, \emph{etc.}~(any additional mechanism for angular-momentum evolution would do). This is essentially equivalent to demanding that the eccentricity change over one outer orbit be of the same order of magnitude as that due to gravitational wave, that is
\begin{equation}
	\frac{1}{T_{\rm out}}\sqrt{\left\langle \Delta e^2 \right \rangle} \sim \left(\frac{\mathrm{d}e}{\mathrm{d}t}\right)_{\rm ext},
\end{equation}
where both sides are evaluated at $e=e_{\rm u}$, and it is assumed that by then $i \approx \pi/2$. Here, $\left(\mathrm{d}e/\mathrm{d}t\right)_{\rm ext}$ is the eccentricity change due to the additional physical mechanism; henceforth we take this to be gravitational-wave emission as an illustrative example, whence in that case
\begin{equation}
	1-e_{\rm u,gw}^2 \geq \left(\frac{68\sqrt{2}T_{\rm out}}{477 \pi \eps \tau_{\rm c}}\right)^{1/3},
\end{equation} 
where $\tau_{\rm c}$ is the time-to-coalescence of a binary, given by
\begin{equation}
	\tau_{\rm c} = \frac{5}{256}\frac{c^5 a_{\rm in}^4}{Gm_{\rm c}^{5/3}m_{\rm b}^{4/3}},
\end{equation}
where $m_{\rm c}$ is the inner binary's chirp mass. 

$e_{\rm u}$ obtained this way is very close to $1$; and we combine this with the requirement in inequality \eqref{eqn:emax from random walk}.
Together
\begin{equation}\label{eqn:e max}
	e_{\rm u,gw}^2 = 1-\max \set{\frac{225\pi^2}{128}\eps,\left(\frac{68\sqrt{2}T_{\rm out}}{477 \pi \eps \tau_{\rm c}}\right)^{1/3}},
\end{equation}
where it is understood that $e_{\rm u} \equiv 0$ if the right-hand side evaluates to a negative number.

\section{First-passage time}
\label{appendix: first passage}
As discussed in \S \ref{sec:external}, the long-term evolution under equations \eqref{eqn: changes per encounter} and \eqref{eqn: evolution} suggests that $e$ and $i$ evolve as a random walk, where $\omega$ and $\Omega$ are essentially viewed as randomised in each step. The continuum limit of this system (where $\eps \to 0$ and the number of steps tends to infinity) is a Fokker--Planck equation, of the form
\begin{equation}\label{eqn:diffusion problem}
	\frac{\partial p}{\partial t} = \frac{\partial}{\partial x^k}\left[D_2^{jk}(e,i)\frac{\partial p}{\partial x^j}\right] + \beta(e,i) \frac{\partial}{\partial x^k}\left[D_1^k(e,i) p\right] \equiv \mathcal{D}[p]\;,
\end{equation}
where $x^j \equiv (e,i)$ is a two-dimensional shorthand, and the diffusion coefficients are $D_1^k, D_2^{jk} = \mathcal{O}(\eps^2)$, because the drift $\langle \Delta e \rangle$ vanishes at first order in $\eps$. The initial condition is 
\begin{equation}
	p(e,i,t=0) = \delta^{\rm D}(e-e_0) \delta^{\rm D}(i-i_0)\;,
\end{equation}
where $\delta^{\rm D}$ is Dirac's delta-function.

For such a process the mean first-passage time, defined as the mean time for a walker to reach $e = e_{\rm u}$, starting at $x_0 = (e_0,i_0)$, is necessarily \citep{Redner2001}
\begin{equation}
	t_{\max}(x_0|e_{\rm u}) \propto \frac{1}{\eps^{2}}\;.
\end{equation}
This may be seen from the associated continuum problem \eqref{eqn:diffusion problem}, where $t_{\max}$ satisfies a backward Kolmogorov equation \citep{Redner2001}
\begin{equation}
	\mathcal{D}^\dagger \left[t_{\max}\right] = -1,
\end{equation}
with a boundary condition $t_{\max}(e_{\rm u},i_0|e_{\rm u}) = 0$. The adjoint Fokker--Planck operator is 
\begin{equation}
	\mathcal{D}^\dagger[t] = \frac{\partial}{\partial x^k_0}\left[D^{jk}(e_0,i_0) \frac{\partial t}{\partial x_0^j}\right] -\beta(e_0,i_0) D_1^{k}(e_0,i_0)\frac{\partial t}{\partial x_0^k}\;,
\end{equation}
and is therefore $\mathcal{D}^\dagger \propto \eps^2$; that is, re-scaling the unit of time $T_{\rm out}$ and $\eps^2$ are interchangeable. This yields equation~\eqref{eqn:diffusion time-scale}.


\bsp	
\label{lastpage}
\end{document}